\documentclass[11pt]{article}
\usepackage{amssymb}\usepackage{amsmath}
\usepackage[all]{xy}
\usepackage[dvips]{graphicx}
\usepackage{pstricks}
\usepackage{pst-plot}

\numberwithin{equation}{section}

\def\be{\begin{equation}}
\def\ee{\end{equation}}
\def\ba{\begin{align}}
\def\ea{\end{align}}

\def\yboxit#1#2{\vbox{\hrule height #1 \hbox{\vrule width #1
\vbox{#2}\vrule width #1 }\hrule height #1 }}
\def\fillbox#1{\hbox to #1{\vbox to #1{\vfil}\hfil}}
\def\ybox{{\lower 1.3pt \yboxit{0.4pt}{\fillbox{8pt}}\hskip-0.2pt}}
\def\comments#1{}

\def\half{\frac{1}{ 2}}
\def\Tr{{{\rm Tr~ }}}

\def\ket#1{|#1\rangle}

\def\II{\relax{I\kern-.10em I}}

\def\IZ{\relax\ifmmode\mathchoice
{\hbox{\cmss Z\kern-.4em Z}}{\hbox{\cmss Z\kern-.4em Z}}
{\lower.9pt\hbox{\cmsss Z\kern-.4em Z}}
{\lower1.2pt\hbox{\cmsss Z\kern-.4em Z}}\else{\cmss Z\kern-.4em
Z}\fi}
\def\IB{\relax{\rm I\kern-.18em B}}
\def\IC{{\relax\hbox{$\inbar\kern-.3em{\rm C}$}}}
\def\ID{\relax{\rm I\kern-.18em D}}
\def\IE{\relax{\rm I\kern-.18em E}}
\def\IF{\relax{\rm I\kern-.18em F}}
\def\IG{\relax\hbox{$\inbar\kern-.3em{\rm G}$}}
\def\IGa{\relax\hbox{${\rm I}\kern-.18em\Gamma$}}
\def\IH{\relax{\rm I\kern-.18em H}}
\def\II{\relax{\rm I\kern-.18em I}}
\def\IK{\relax{\rm I\kern-.18em K}}
\def\IN{\relax{\rm I\kern-.18em N}}
\def\IP{\relax{\rm I\kern-.18em P}}

\def\inbar{\,\vrule height1.5ex width.4pt depth0pt}

\font\cmss=cmss10 \font\cmsss=cmss10 at 7pt
\def\IR{\relax{\rm I\kern-.18em R}}

\def\lp10{l_P^{10}}
\def\lp11{l_P^{11}}
\def\R11{R_{11}}

\def\Tr{{\rm Tr}}

\font\manual=manfnt
\def\dbend{\lower3.5pt\hbox{\manual\char127}}

\def\IZ{\relax\ifmmode\mathchoice
{\hbox{\cmss Z\kern-.4em Z}}{\hbox{\cmss Z\kern-.4em Z}}
{\lower.9pt\hbox{\cmsss Z\kern-.4em Z}} {\lower1.2pt\hbox{\cmsss
Z\kern-.4em Z}}\else{\cmss Z\kern-.4em Z}\fi}
\def\half {\frac{1}{ 2}}

\def\bar{\overline}

\def\rt2{\sqrt{2}}
\def\irt2{\frac{1}{\sqrt{2}}}

\def\b{\beta}
\def\a{\alpha}

\font\cmss=cmss10
\font\cmsss=cmss10 at 7pt
\def\IL{\relax{\rm I\kern-.18em L}}
\def\IH{\relax{\rm I\kern-.18em H}}
\def\IR{\relax{\rm I\kern-.18em R}}
\def\inbar{\vrule height1.5ex width.4pt depth0pt}
\def\IC{\relax\hbox{$\inbar\kern-.3em{\rm C}$}}
\def\rlx{\relax\leavevmode}
\def\ZZ{\rlx\leavevmode\ifmmode\mathchoice{\hbox{\cmss Z\kern-.4em Z}}
 {\hbox{\cmss Z\kern-.4em Z}}{\lower.9pt\hbox{\cmsss Z\kern-.36em Z}}
 {\lower1.2pt\hbox{\cmsss Z\kern-.36em Z}}\else{\cmss Z\kern-.4em
 Z}\fi}
\def\IZ{\relax\ifmmode\mathchoice
{\hbox{\cmss Z\kern-.4em Z}}{\hbox{\cmss Z\kern-.4em Z}}
{\lower.9pt\hbox{\cmsss Z\kern-.4em Z}}
{\lower1.2pt\hbox{\cmsss Z\kern-.4em Z}}\else{\cmss Z\kern-.4em
Z}\fi}

\def\Tr{{\rm Tr}}

\font\manual=manfnt
\def\dbend{\lower3.5pt\hbox{\manual\char127}}

\def\IZ{\relax\ifmmode\mathchoice
{\hbox{\cmss Z\kern-.4em Z}}{\hbox{\cmss Z\kern-.4em Z}}
{\lower.9pt\hbox{\cmsss Z\kern-.4em Z}} {\lower1.2pt\hbox{\cmsss
Z\kern-.4em Z}}\else{\cmss Z\kern-.4em Z}\fi}
\def\half {\frac{1}{ 2}}

\def\bar{\overline}

\def\rt2{\sqrt{2}}
\def\irt2{\frac{1}{\sqrt{2}}}

\input{epsf}

\usepackage{epsfig}
  
\title{\Large{\bf Non-compact Gepner Models, Landau-Ginzburg Orbifolds and
    Mirror Symmetry}} \author{Sujay K. Ashok$^{a,b}$, Raphael Benichou$^{c}$
  and Jan Troost$^{c}$ } \date{}
\begin{document}
\maketitle

\begin{center}
  $^{a}$Institute of Mathematical Sciences\\
  C.I.T Campus, Taramani\\
  Chennai, India 600113\\
  $^{b}$Perimeter Institute for Theoretical Physics\\
  Waterloo, Ontario, ON N$2$L$2$Y$5$, Canada \\
  $^{c}$Laboratoire de Physique Th\'eorique\footnote{Unit\'e Mixte du CRNS et
    de l'Ecole Normale Sup\'erieure associ\'ee \`a l'universit\'e Pierre et
    Marie Curie 6, UMR
    8549. Preprint LPTENS-07/51.}, Ecole Normale Sup\'erieure \\
  $24$ Rue Lhomond Paris $75005$, France
\end{center}

\begin{abstract}
  We study non-compact Gepner models that preserve sixteen 
or eight
  supercharges in type II string theories. In particular, we develop an
  orbifolded Landau-Ginzburg description of these models analogous to the
  Landau-Ginzburg formulation of compact Gepner models. The Landau-Ginzburg
  description provides an easy and direct access to the geometry of the
  singularity associated to the non-compact Gepner models.  Using these tools,
  we are able to give an intuitive account of the chiral rings of the models,
  and of the massless moduli in particular. By studying orbifolds of the
  singular linear dilaton models, we describe mirror pairs of non-compact
  Gepner models by suitably adapting the Greene-Plesser
  construction of 
mirror pairs for the compact case. For particular models, we
  take a large level, low curvature limit in which we can
  analyze corrections to a flat space orbifold approximation of the
  non-compact Gepner models. This gives rise to a counting of moduli which
  differs from the toric counting in a subtle way.
\end{abstract}

\section{Introduction}
Mirror symmetry for Calabi-Yau $3$-folds is a subject of great interest 
to physicists as well as mathematicians \cite{Yau:1998yt}\cite{Greene:1997ty}.
Mirror pairs were first exhibited \cite{Greene:1990ud}
by studying orbifolds of the quintic at the Gepner point \cite{Gepner:1987qi}
in the Calabi-Yau moduli space. This was explored in more detail in
\cite{Candelas:1990rm} which contained the first mathematical predictions from
mirror symmetry. Many generalizations were found using the Landau-Ginzburg formulation that were applicable in non-geometric regimes \cite{Berglund:1991pp}.

In this paper, we wish to extend the study of mirror symmetry to non-compact
Gepner models. Various approaches towards this problem have been followed.
Toric Calabi-Yau manifolds have been very well studied in the literature
following the construction of mirror pairs for hypersurfaces in toric
varieties \cite{Batyrev} and for non-compact Calabi-Yau manifolds (see
\cite{Chiang:1999tz} and references therein). In the physics literature, this
has been reformulated using a gauged linear sigma model \cite{Witten:1993yc,
  Hori:2000kt, Hori:2002cd}. Progress has also been made for more general
$c=9$ theories via conformal field theory techniques \cite{Eguchi:2004yi}. In
particular, the work on T-duality for the $N=2$ cigar and the Liouville
conformal field theories is relevant in this context \cite{Giveon:1999px,
  Hori:2001ax, Tong:2003ik, Israel:2004jt}.

One of the gaps we aim to fill in this paper is to clarify the link between
non-compact Gepner models and their Landau-Ginzburg description. For compact
Calabi-Yau manifolds that are hypersurfaces in weighted projective space, the
Landau-Ginzburg description is closely tied to the geometry
\cite{Greene:1988ut}. These are two of the many phases in a gauged linear
sigma-model description \cite{Witten:1993yc}. The Landau-Ginzburg model is also extremely useful for providing a simple setting to do computations. For instance, it has served well in the
past for establishing the link between compact Gepner models and compact
Calabi-Yau manifolds \cite{Greene:1988ut} as well as to provide the first list
of Calabi-Yau hypersurfaces in $W\IC\IP^4$ \cite{Candelas:1989hd}.

Our approach in this paper is to stress the ingredients of the analysis
of Gepner models that survive the transition from going from the compact to
the non-compact case. We alo point out the characteristics that differ
in the two cases.
Moreover we study in detail the orbifolded Landau-Ginzburg description of these models
\cite{Vafa:1989xc, Intriligator:1990ua}. This allows us to 
compare our models with geometric backgrounds which can be used as
internal spaces for type II strings on $\IR^{3,1}$.

In section \ref{noncompactgepner} we will review the asymptotic
partition function of non-compact Gepner models \cite{Eguchi:2000tc,Hanany:2002ev,Murthy:2003es,Eguchi:2004yi,Israel:2004ir}. The asymptotic partition
function, which is proportional to the divergent volume of space-time,
otherwise behaves much as
the partition function in the compact case. In particular we show here that
the $\beta$-method of Gepner \cite{Gepner:1987qi} for constructing modular
invariants can be adapted to the non-compact case. We then start the
discussion of the localized spectrum
\cite{Hanany:2002ev,Eguchi:2004yi,Israel:2004ir}, which is
 characteristic of non-compact models.

 In section \ref{LG} we will link the non-compact Gepner models to
 Landau-Ginzburg models. We briefly remark on the difference between the
 compact and non-compact Landau-Ginzburg model \cite{Ooguri:1995wj}. For
 instance, the former gives rise to a unital chiral ring, while the latter
 gives rise to a ring without unit element. We continue the analysis in
 section \ref{orbifoldedLG} with a discussion of the orbifolds of the
 Landau-Ginzburg models that are necessary to implement the GSO projection,
 and other orbifold groups. We will see that the formalism for counting chiral
 ring elements largely carries over from the compact case
 \cite{Vafa:1989xc,Intriligator:1990ua}. However, it will become intuitive
 that the Landau-Ginzburg potentials with negative powers exclude some of the
 (anti-)chiral ring elements as the potential renders them non-normalizable.
 Continuing the formal counting exercise will turn out to be useful
 nevertheless. It gives rise to a natural picture of mirror symmetry in
 conformal field theories, that is strongly reminiscent of its compact
 counterpart \cite{Greene:1990ud, Eguchi:2004yi}. We will discuss this point
 in section \ref{CFTmirror}.

 In section \ref{examples} we provide new examples of mirror conformal field
 theories. The techniques developed to analyze orbifolded Landau-Ginzburg
 models come in handy when treating these more complicated models. The mirror
 pairs of conformal field theories are best understood as linear dilaton
 backgrounds with $N=2$ superconformal symmetry. These 
can be deformed or resolved, to give rise to perturbatively well-defined mirror string backgrounds.  Finally, in section \ref{geometrymirror}, we discuss examples in which we can relate the mirror conformal field theories to geometries. It will turn out that we can approximate certain conformal field theories at large level with orbifold singularities that admit a toric description. At infinite level, we find agreement between the conformal field theory and the geometric results. At finite level, we find that the conformal field theory description takes into account various modifications to the background that can lift some moduli. In section \ref{conclusions} we discuss further applications of our results.

\section{Non-compact Gepner Models}
\label{noncompactgepner}
We start out with a rather brief but technical review of non-compact Gepner models (see
e.g. \cite{Eguchi:2000tc,Murthy:2003es,Eguchi:2004yi,Israel:2004ir})
in order to clarify the fact that the asymptotic partition function of
non-compact Gepner models can be constructed using the same
tools as in the compact case. 
We can then lay bare properties of the models along the same lines as in the
compact Gepner models.  In the second part of this section we give a preview
of the ingredients that go into analyzing the localized part of the partition
function.

\subsection{The Asymptotic Non-compact Gepner Models}
 \label{asymptotic}

For simplicity we will work with type II string theory on $\IR^{3,1}$ times an
internal (non-compact) conformal field theory of central charge
nine\footnote{The constructions can be extended to lower-dimensional flat
  spaces and to heterotic string theories, with little effort and lots of
  indices.}.  The internal conformal field theory is built from a product of
$N=2$ superconformal field theories. These can be split into three classes
depending on whether their central charge is smaller, larger 
than or equal to
three.

\begin{itemize}
\item[\_]The minimal $N=2$ superconformal field theories  (see
  e.g. \cite{Gepner:1986hr})
have central
charge smaller than three. They can be
viewed as coset conformal field theories of the form $SU(2)_{k-2} \times
U(1)_2 / U(1)_k$ of central charge $c_{MM}=3 -\frac{6}{k}$. The level $k$ is
the supersymmetric level of the total $SU(2)$ current algebra present in the
parent $N=1$ Wess-Zumino-Witten model. It is a positive integer
larger than or equal to two.

 Since the minimal $N=2$ superconformal models
have been reviewed frequently 
(\cite{Greene:1996cy,Maldacena:2001ky}), and since they are
standard in the construction of Gepner models \cite{Gepner:1987qi}, we only
very briefly recall some of their properties. The primaries of the model can
be labeled by three quantum numbers: the spin $j$ under the $SU(2)_{k-2}$
current algebra, the $\IZ_{2k}$ valued chiral momentum $n$ under the $U(1)_k$
current algebra and the $\IZ_4$ valued chiral momentum $s$ labeling a
representation of the $U(1)_2$ current algebra.  They satisfy the selection
rule: $2j\equiv n+s\ [2]$. We moreover have an equivalence between the
following representations: $(j,n,s) \equiv (\frac{k-2}{2}-j,n-k,s+2)$.  The
left-moving $U(1)_R$ charge $Q_{MM}$ for a primary is equal to $Q_{MM}=
\frac{n}{k} + \frac{s}{2}$.

\item[\_]The second class of theories has central charge
larger than three. An example in this class is an $N=2$ linear dilaton theory
with a slope such that the central charge is equal to $c=3+\frac{6}{l}$ with
positive and real values for the parameter $l$.
The superconformal algebra with central charge larger than three has
continuous unitary representations which are conveniently labeled
\cite{Dixon:1989cg} by a Casimir $j=1/2 + ip $ where $p \in \IR^+$, by an
integer momentum $2m \in \IZ$ and a $\IZ_4$ fermionic quantum number $s$. The
left-moving R-charge $Q_{nc}$ of a primary is $Q_{nc} = \frac{2m}{l}
+\frac{s}{2}$. (See e.g. \cite{Israel:2004jt} for a detailed discussion.)

\item[\_]The $N=2$ superconformal
algebra with $c=3$ is exceptional, and can be represented for instance by free
scalars (which can realize compact or non-compact target space directions).
\end{itemize}

In order to mimic the Gepner construction, we will assume that in the case
$c>3$, the chiral
algebra has some more structure (see e.g. \cite{Eguchi:2000tc}). For simplicity we will
work under the assumption that the parameter $l$ is a positive integer\footnote{It is sufficient to suppose that the central
charge is parameterized by a positive fractional level $l$ \cite{Eguchi:2000tc}.}.
We can then add to
the chiral algebra the generator of spectral flow on the $N=2$ superconformal
algebra by $2 l$ units.  The characters of the extended $N=2$ superconformal
algebra in the continuous representations are given by:
\begin{eqnarray}
Ch_{cont} (j,2m,s) &=& q^{\frac{p^2}{l}}   \frac{1}{\eta^3(\tau)}
\Theta_{s,2}(\tau) 
\Theta_{2m,l} (\tau).
\end{eqnarray}
It is crucial to us that the modular transformation properties of the
characters hinge upon the presence of the  
$\theta$-functions at levels $2$ and $l$.

\subsubsection{Levels And Charge Lattice}

We observe that the characters of the $N=2$ minimal models at
level $k$ transform as $\theta$-functions at levels $2$ and $-k$,
while the extended characters of the $N=2$ models with central charge
$c=3+\frac{6}{l}$ transform as $\theta$-functions at level $2$ and
$+l$. We thus note a first important sign difference 
in the transformation rules of the characters. Modular
invariants in the continuous sector of the theory can  be based on
modular invariants of $\theta$-functions. For one $U(1)_k$ current algebra
at level $k$
these are well-known to correspond to the divisors of $k$ via
orbifolding of the diagonal modular invariant. For a product of
$\theta$-functions, the analysis is more complicated,
 but a large class of modular invariants can be constructed by
taking products of modular invariants of the factors, and then
orbifolding.

In order to write down the modular invariant partition functions,
it is useful to introduce a charge lattice for the various $U(1)$
current algebras in
the theory. We introduce a vector of levels 
$(2,2, \dots, 2; k_1, \dots , k_p ; l_1, \dots, l_q)$ 
where $p$ is the number of
minimal model factors, $q$ is the number of non-compact factors, while
the number of fermionic levels equal to $2$ is $1+q+p$. Indeed, we work
 in light-cone gauge on $\IR^{3,1}$ such that there is one complex fermion
associated to the flat space directions (and there is
one complex fermion per factor
model). 
The charge lattice is periodic. In each direction, the periodicity is twice
the level.
A point in the lattice is defined by a charge vector $r= (s_0,s_1, \dots,
s_{p+q}; n_1,\dots,n_p; 2m_1, \dots, 2m_q)$ where we used the
notation
$s_0$ for the charge of the flat
space fermions under the $U(1)_2$ current algebra, and
similarly for the other fermions, while we copy 
the traditional notation for the
chiral momentum quantum numbers of compact and non-compact factors
that we introduced above (including their normalization). 
The scalar product on the charge lattice is defined as follows:
\begin{eqnarray}
r^{(1)} \cdot r^{(2)} &=& -\frac{s_0^{(1)} s_0^{(2)}}{4} - \frac{s_1^{(1)} s_1^{(2)}}{4} - \dots
- \frac{s_{p+q}^{(1)} s_{p+q}^{(2)}}{4} 
\nonumber \\ 
& & + \frac{n_1^{(1)} n_1^{(2)}}{2k_1} + \dots  + \frac{n_p^{(1)} n_p^{(2)}}{2k_p}
\nonumber \\
& & - \frac{(2m_1^{(1)})(2m_1^{(2)})}{2l_1} -\dots - \frac{(2m_q^{(1)})(2m_q^{(2)})}{2l_q}\,.                               
\end{eqnarray}
The contribution of the chiral momenta corresponding to the non-compact factors comes with an opposite sign from those of the compact factors. The all-important signature of the
quadratic form is therefore $(-,\dots,-; +, \dots, + ; - , \dots, -)$.

\subsubsection{A Canonical Vector}

We introduce the following vector $\beta_0$ in the charge lattice\footnote{We
  take the entries for the fermions to be minus one, in order to accord with
  the convention that the left-moving $U(1)_R$ charge 
is given for instance for a minimal
  model factor by $Q_{MM}=\frac{n}{k}+\frac{s}{2}$.}:
\begin{eqnarray}
\beta_0 &=& (-1,-1, \dots, -1; 1,1,\dots,1; -1,-1, \dots,-1).
\end{eqnarray}
We have that the left-moving R-charge $Q$ for a primary state with charge
vector $r$ is equal to $Q=2 \beta_0 \cdot r$. We moreover have that
\begin{eqnarray}
\beta_0 \cdot \beta_0 &=& -\frac{1+p+q}{4} + \sum_{i=1}^p \frac{1}{2k_i} - \sum_{j=1}^q \frac{1}{2l_i}
\nonumber \\
&=& -1
\end{eqnarray}
where we used that the total central charge of the light-cone gauge conformal field theory is equal to 
\be
12 = 3(1+p+q) -  \sum_{i=1}^p \frac{6}{k_i} + \sum_{j=1}^q \frac{6}{l_i} \,.
\ee
In summary, the vector $\beta_0$ is useful to measure the R-charge, and
squares to minus one.

For the right-movers we will always take identical conventions to the
left-movers. In particular, the $N=2$ superconformal algebras have the
same structure constants. We will also define the right-moving
R-charge of the minimal model factors to be $\tilde{Q}_{MM} =
\frac{\tilde{n}}{k} + \frac{\tilde{s}}{2}$ and for the non-compact factors
$\tilde{Q}_{nc}=\frac{2 \tilde{m}}{k} + \frac{\tilde{s}}{2}$, while the charge vector
for the right-movers is $\tilde{r} = ( \tilde{s}_0, \dots, ; \dots ;
\dots, 2 \tilde{m}_{p+q})$. So, for the right-movers as well we have
that the total $U(1)_R$ charge is given by $\tilde{Q}= 2 \beta_0 \cdot
\tilde{r}$ with the same vector $\beta_0$.

\subsubsection{Products of $\theta$-functions}
We define the following notation for the product of characters of the
 flat space fermions, the minimal model and the non-minimal $N=2$
 superconformal field theories. Since the characters transform like
 $\theta$-functions, we introduce the symbol:
\begin{eqnarray}
\Theta_{r} (\tau) &=& \Theta_{s_0,2}(\tau) \prod_{i=1}^p \chi_{j_i,n_i,s_i} (\tau)
 \prod_{i=1}^q Ch_{cont}(j_{p+i},2m_i,s_{p+i}) (\tau)
\end{eqnarray}
where $r$ is the total charge vector, and the first factor corresponds
to the flat space fermions, while the following $p$ factors correspond to
minimal model characters, 
and the final $q$ factors to the non-compact continuous
extended characters. In the $\Theta$ symbol we have left implicit the
labels corresponding to the levels, as well as those corresponding to
the Casimirs of the minimal and non-minimal models. The important point is that
the $\Theta$-functions transform as a product of ordinary $\theta$-functions
under modular transformations.

We introduce now the first modular invariant partition function which is the
diagonal modular invariant:
\begin{eqnarray}
Z_{diag} &=& \sum_{r} \Theta_{r} (\tau) \Theta_r (\bar{\tau})\,, 
\end{eqnarray}
where the diagonal sum over $r$ is over all inequivalent charges in the charge
lattice.  Implicitly, we take a diagonal A-type modular invariant for the
Casimir invariants $j$ for all factors\footnote{It would be interesting to
  study the more general modular invariants of type D and/or type E for the
  compact factors.}. We have suppressed the divergent non-compact volume factor
in the formula for the asymptotic partition function.

\subsubsection{Locality Orbifold}

Now that we have set-up our theory in a way which is very analogous to
\cite{Gepner:1987qi}, we can follow that reference closely. Along the way, we
reformulate a few minor points in a more modern orbifold language.

Locality in string theory 
only allows for fermions having either
Neveu-Schwarz (NS) or Ramond (R) boundary conditions for all factors simultaneously for the
left- or the right-movers. We implement that locality constraint by
orbifolding the diagonal partition function by a diagonal $(\IZ_2)^{p+q}$ group.
The i'th $\IZ_2$ acts on a state with charge vectors $r$ and $\tilde{r}$ as
$(-1)^{(s_0+s_i+\tilde{s}_0 + \tilde{s}_i)/2}$ for $i=1, \dots, p+q$. The action
can be summarized by introducing vectors $\beta_i$ which have a $2$ as
the first and the $i+1$'th entry for $i=1, \dots, p+q$. 
Then the action can equivalently be written as $(-1)^{\beta_i \cdot (r+\tilde{r})}$.
On an untwisted state with charges $s_0$ and $s_i$, the action is the
multiplication by a phase $(-1)^{s_0+s_i}$. 

The states invariant under $(\IZ_2)^{p+q}$ will be purely NS or purely R for
the left-movers. Since the partition function was diagonal, the same condition
will hold for the right-movers.  The orbifold also introduces twisted states.
These have left- and right-moving charge vectors that differ by multiples of
the vectors $\beta_i$ \cite{Gepner:1987qi}. Indeed, since the fermion number
$s_i$ is defined modulo four, this introduces a twisted state sector for each
$\IZ_2$ orbifold factor, and moreover, since $\beta_i^2=2$, we have that the
twisted states we introduce in this fashion are also orbifold invariant
\cite{Gepner:1987qi}.  Each of the $\IZ_2$ orbifolds introduces twisted
states, which renders the sum over left- and right-moving fermion numbers $s_i$
and $\tilde{s_i}$ independent, except for the fact that they need to be of the
same parity as the flat space fermion numbers $s_0$ and $\tilde{s}_0$
respectively. The above prescription is equivalent to the standard orbifold
procedure and produces a new modular invariant partition function.  Note that
the flat space fermion quantum numbers $s_0$ and $\tilde{s}_0$ are still of
equal parity.  We therefore only have NSNS and RR states at this point.  The
sum over the left- and right-moving worldsheet fermion numbers $s_0$ and
$\tilde{s}_0$ will become decoupled after performing a final $\IZ_2$ GSO
projection, thus introducing fermions.

\subsubsection{Integer R-charge Orbifold}
\label{integerR}

The standard GSO projection in string theory is based on the fact that
the partition function only has integer R-charges. In order to ensure
this condition in a Gepner model, we perform yet another orbifold. The
orbifold action on a state with charge $r$ and $\tilde{r}$ will be $exp
(2 \pi i \beta_0 \cdot (r + \tilde{r}))$. We first note that $\beta_0
\cdot \beta_i$ is an integer, such that the action of the orbifold on
$\beta_i$ twisted sectors is identical to the action on  $\beta_i$ untwisted
sectors. The order of the $2 \beta_0$ orbifold is therefore the order
of the operator $e^{ 2 \pi i Q } $ in the untwisted theory (where
$Q$ is the total left-moving R-charge in light-cone gauge)\footnote{The theory
we start out with obeys the charge relation $r=\tilde{r}$. 
Since $\beta_0 \cdot \beta_i \in \IZ$, we
still have the relation $e^{ 2 \pi i (Q+\tilde{Q})/2 } = e^{ 2 \pi i Q } $ after the
$(\IZ^{p+q}_2)$ orbifold.}. The order of
that orbifold is the smallest common divisor $d$ of all the levels in the
theory (including the fermionic level $2$ when $p+q$ is even and 
not when $p+q$ is odd).

It is clear in the untwisted sector that the orbifold forces the
 left-moving (and right-moving) $U(1)_R$ charge to be integer. There
 are also $d-1$ twisted sectors which have charge vectors which differ
 by multiples of $ 2 \beta_0$. Since $\beta_0^2=-1$, we have that
 these twisted sectors also have integer left- and right-moving
 R-charges. Thus, the orbifold has provided us with a new modular
 invariant partition function with integer R-charges for both left-
 and right-movers.  If we introduce the lattice $\Lambda$ generated
by the vectors $\beta_i$ and $2 \beta_0$, then we can write the
 partition function of the theory
 as\footnote{This partition function for non-compact models
is the analogue of $Z_0$ in
 \cite{Greene:1990ud} for compact Gepner models.}:
\begin{eqnarray}
Z_0 &=& \sum_{r-\tilde{r} \in \Lambda} \Theta_r (\tau) \Theta_{\tilde{r}} (\bar{\tau})\,,
\end{eqnarray}
where we restrict the sum to invariant states, namely states obeying the
conditions 
$ r \cdot \beta_i \in \IZ$ (purely NS or purely R) and 
$ r \cdot 2 \beta_0 \in \IZ$ (integer R-charges). Again we have left implicit the diagonal
A-type invariant for the minimal models as well as the diagonal integral over
radial momenta for the compact factors that diverges like the volume of space-time.

\subsubsection{The Standard GSO Projection}

The partition function now has integer R-charges for all states, and can be
GSO projected in the same manner as the flat space partition function. We
project onto odd R-charges, i.e. we satisfy the condition $2 \beta_0 \cdot r
\in 2 \IZ +1$ as well as $2 \beta_0 \cdot \tilde{r} \in 2 \IZ+1$.
The charge difference between left- and right-movers for the twisted states is
proportional to an odd multiple of $\beta_0$. The twisted sectors are the NS-R
and R-NS sectors of the theory, which correspond to space-time fermions, and
contribute negatively to the space-time partition function thus implementing
space-time statistics.  For type II theories, we have obtained an
asymptotic supersymmetric partition function in the Gepner formalism. Strictly
speaking we have a type IIB partition function, since we have made no
distinction between left- and right-movers.  It can be transformed easily into
a type IIA partition function by flipping the chirality of the final
GSO projection for the right-movers in the R-sector.

\subsubsection{Discrete Symmetries}
\label{symGroup}
 In this subsection, we can follow  \cite{Greene:1990ud} closely
since we have set up our model as in the compact case.
It is interesting to single out a particular symmetry group of the
model. 

The symmetry of the compact models contains a $\IZ_{k_i} \times \IZ_{k_i}$
group. Only the diagonal $\IZ_{k_i}$ subgroup has a non-trivial action on a
model with diagonal spectrum.
This subgroup acts as 
\be
\Phi_{r,\tilde{r}} \rightarrow exp(2 \pi i
(n_i+\tilde{n}_i)/ 2 k_i) \Phi_{r,\tilde{r}}\,.
\ee 
We can introduce a vector
\be 
\gamma_i=(0,...,0;0,...,0,2,0,...,0; 0,...,0)
\ee 
where $2$ is in the $i$'th entry after the first semi-column. The vector codes the action of the symmetry
group as follows: 
\be
\Phi_{r,\tilde{r}} \rightarrow exp(\pi i \gamma_i \cdot (r +
\tilde{r}))\Phi_{r,\tilde{r}}\,.
\ee 
We have a similar action in the non-compact theories. There is 
a symmetry group $\IZ_{l_i} \times\IZ_{l_i}$, of which only the diagonal subgroup acts nontrivially on the
states. In the full Gepner model, we can think of the product of the diagonal
subgroups 
\be
D=\prod_{i=1}^p \IZ_{k_i} \times \prod_{j=1}^q \IZ_{l_j}
\ee 
as mapped into the charge lattice via the maps $\gamma_{i=1, \dots, p+q}$.

Now we define the operator $g_0$ that acts by multiplication by $\exp(2\pi i
\beta_0 \cdot(r+\tilde{r}))$. It generates a group $\IZ_d$, that we used
previously to perform the integer R-charge orbifold.
This orbifold group $\IZ_d$ contains a subgroup
generated by $g_0^2$, which is a subgroup of
$D$. Indeed, the
element $g_0^2$ corresponds to a $\beta$-vector that has zero fermionic
entries.  When the order $d$ of the group $\IZ_d$ is even, the subgroup
generated by $g_0^2$ has order $n=d/2$ and it has order $n=d$ when $d$ is odd
(see also \cite{Greene:1990ud}).  The part of the diagonal symmetry group $D$ that
still acts after the projection onto integer R-charges is 
\be
G=(\prod_{i=1}^p
\IZ_{k_i} \times \prod_{j=1}^q \IZ_{l_j})/\IZ_n\,.
\ee

Following \cite{Greene:1990ud} we then define the maximal subgroup $H$ of $G$
which preserves supersymmetry in space-time. This is the subgroup
corresponding to all vectors $\beta_m$ in the charge lattice that satisfy the
equation
$2
\beta_m \cdot \beta_0 \in 2 \IZ$. 
This condition ensures that the left- and right- moving R-charges only differ
by even integers, as required by supersymmetry.
If we write 
\be
\beta_m = \sum_i c_m^i \gamma_i\,,
\ee 
then this condition boils down to the condition that \cite{Greene:1990ud}
\be
\sum \frac{c_m^i}{k_i} + \sum \frac{c_m^j}{l_j} \in \IZ \,.
\ee 
The subgroup $H$ generated by the vectors $\beta_m$ that satisfy this condition is the
maximal orbifold group consistent with space-time supersymmetry.

Any subgroup $F$ of $H$ can be used to generate new supersymmetric orbifold
models. Precisely as in the compact case, modding out the original
model by the maximal subgroup $H$ generates the mirror model
(in the conformal field theory sense). Moreover, orbifolding by a subgroup $F$
will generate a model that is mirror to the orbifold of the original model by
$H/F$. This was argued in \cite{Greene:1990ud} for the compact case, and we
will show in section \ref{CFTmirror} that it is also true for the non-compact
models. Note that the mirror symmetry in the
conformal field theory that we have set up above holds for {\em
  undeformed} models. We will see that deformations of a model are mapped to
resolutions of the mirror. We will discuss this important point in greater detail
later on.

\subsection{The Deep-throat Region Of Non-compact Gepner Models}
\label{local}
Until now we have discussed only the continuous part of the spectrum of
the non-compact Gepner models. 
The contribution of these states to the partition function is proportional to
the volume of the target space.

A good starting point for the rest of our discussion will be to think
of the initial model as based on a product of $N=2$ superconformal
minimal models and $N=2$ linear dilaton theories.

It is important to observe that although the exact torus partition
function exists for these conformal field theories, 
they only describe the
asymptotic spectrum of the corresponding string 
theory. Indeed, there is a region in space-time that is strongly
coupled (due to the linearly growing dilaton in one or several
directions). In that region, the one-loop spectrum is not a meaningful
quantity.

Nevertheless, we can get a handle on possible deformations of that
singular string theory by working under the following hypotheses. We
look for local deformations in the strongly coupled region deep
in the throat(s).
Secondly, we focus on 
marginal deformations of the worldsheet theory that
preserve supersymmetry in space-time. The $N=2$ superconformal algebra on the
worldsheet will be preserved, and the deformations will be based on chiral (or
anti-chiral) primaries. Morever, we suppose that the deformations cannot have
$U(1)_R$ quantum numbers that differ from those already appearing in the
asymptotic partition function, namely, the charges are quantized as in the
asymptotic partition function.  We believe all of these are mild assumptions,
given space-time supersymmetry. (See e.g. \cite{Gukov:1999ya}
\cite{Shapere:1999xr} for similar reasonings, mostly from a space-time
perspective.)

Using the fact that there is a map between $N=2$
superconformal algebra representations and reprentations of $SL(2,\IR)$
\cite{Dixon:1989cg}, we can 
reformulate the above conditions as saying that the marginal operators in the
full theory should
be based on chiral primaries of dimension smaller or equal to one half in
the linear dilaton factors. The operator in the linear dilaton factor
 should have a conformal dimension equal to $h=|m|/k$ where the $U(1)_R$
charge is given by $Q=2m/k$, and where $2m$ is an allowed $U(1)_R$ quantum
number given the (fixed) asymptotic partition function.
We get an upper cut-off: $2|m| \le k$ from the requirement of relevance on the
worldsheet. 
Moreover, we want to study operators that are
normalizable at weak coupling.
Strict normalisability requires $2 |m|>1$. In particular
the value $2 m=0$ is excluded in the non-compact factors, since this would
correspond to a non-normalizable operator for a conformal field theory with a
non-compact target space.
In summary, the quantum number $2|m|$ has to lie in the range
$1 < 2|m| \le k$. 
It should be noted that the operator with $2|m|=1$ is on the border
of being normalizable in the sense that it lies at the endpoint of a line of
delta-function normalizable operators \cite{Dixon:1989cg}. It will play a special role in what
follows, and we will call this type of operator almost-normalizable. We will flesh out the above analysis considerably in the following sections.

\begin{figure} \centering
\includegraphics{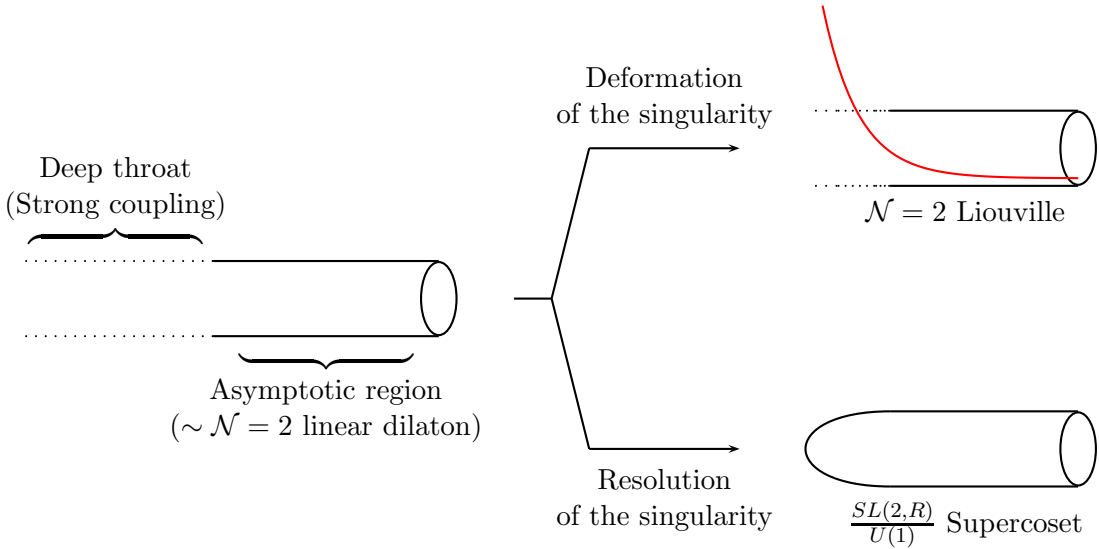}
\caption{
The asymptotic region in a non-compact Gepner  model is identical to that of a linear dilaton conformal field theory. The strong coupling singularity can be cured either by turning on a Liouville potential, or by capping the cylinder with a cigar type deformation. 
}
\label{defres}
\end{figure}

\section{Landau-Ginzburg Models}
\label{LG}
One of our goals in this section is to obtain, in a simple manner, the chiral ring of localized operators in 
the non-compact Gepner models we described in section \ref{noncompactgepner}.
For the compact case, this is most carried out by associating a Landau-Ginzburg model to
each of the factor conformal field theories. The underlying idea is that at low energies,
the Landau-Ginzburg model flows to the conformal field theory that corresponds to the minimal model. 

Furthermore, the GSO projection that we discussed for the non-compact Gepner
models maps to an orbifold of the Landau-Ginzburg theory. Therefore the
techniques derived in \cite{Vafa:1989xc, Intriligator:1990ua} to obtain the
spectrum in the Landau-Ginzburg models and orbifolds thereof will be crucial
for our analysis.

In this section and the next, we give a similar Landau-Ginzburg
description of our non-compact Gepner models. We will find that much of the
technology used in the compact case can be used for the non-compact case
as well. However there are subtle differences in reading off the spectrum,
because some of the operators are not normalizable.

\subsection{Landau-Ginzburg Potentials For Minimal Models}

For $N=2$ superconformal minimal models, the flow between Landau-Ginzburg and
superconformal minimal models is well-studied
\cite{Martinec:1988zu, Vafa:1988uu} and we merely give a
 brief
reminder. A Landau-Ginzburg model with $N=(2,2)$ supersymmetry and a chiral
superfield $\Phi$ with superpotential \be\label{simpleLG} W_{MM} = \Phi^k \ee
flows in the infrared to an $N=2$ superconformal minimal model. The chiral
ring of both models match one-to-one. The chiral unital ring of the Landau-Ginzburg
model is
 $\, \, \IC[\Phi]/\partial_\Phi W$ which is linearly generated by the $k-1$
elements $\Phi^0, \Phi^1, \dots, \Phi^{k-2}$, which have (both left- and
right-moving) R-charge equal to $0,\frac{1}{k}, \dots , \frac{k-2}{k}$. These
match one-to-one to the chiral-chiral primaries of the diagonal minimal model.
Further evidence for this identification of the Landau-Ginzburg model fixed
point is provided by the matching of the elliptic genus of these models
\cite{Witten:1993jg}.

The T-dual or mirror Landau-Ginzburg model based on an twisted chiral
superfield \cite{Rocek:1991ps} flows to the anti-diagonal minimal model. This
can be written as a
$\IZ_k$ orbifold of the model in 
equation \eqref{simpleLG}. We will discuss such orbifolding methods to compute mirrors in later sections.

\begin{figure}
\centering
\includegraphics{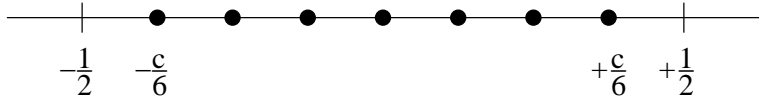}
\caption{The R-charges of the ground states in the Ramond sector
for a $N=2$ superconformal theory with central charge $c<3$.\label{Rgroundstatescomp}}
\end{figure}

We can summarize the chiral ring of the Landau-Ginzburg model by specifying a
Poincare polynomial 
which is the trace over the chiral-chiral ring weighted
by the $U(1)$ R-charges \cite{Lerche:1989uy}:
\begin{eqnarray}
Tr_{(c,c)} t^{Q} \tilde{t}^{\tilde{Q}} &=& 1+ (t \tilde{t})^{\frac{1}{k}} + (t \tilde{t})^{\frac{2}{k}}
+ \dots +  (t \tilde{t})^{\frac{k-2}{k}}
\nonumber \\
& =& \frac{ 1- (t \tilde{t})^{\frac{k-1}{k}} }{ 1 - (t \tilde{t})^{\frac{1}{k}}}.
\end{eqnarray}
In the Ramond sector, this gives rise to a polynomial that keeps track of the
R-charges
of the Ramond-Ramond ground states:
\begin{eqnarray}
Tr_{RR} t^{Q} \tilde{t}^{\tilde{Q}} &=&(t \tilde{t})^{-\frac{1}{2}+\frac{1}{k}} + (t
\tilde{t})^{-\frac{1}{2} +\frac{2}{k}}+ \dots + (t \tilde{t})^{+\frac{1}{2}-\frac{1}{k}}.
\end{eqnarray}
Note that the charges fill out the range from $-c/6$ to $+c/6$, and lie
inside
the interval $]-\half,+\half[$. See figure \ref{Rgroundstatescomp}. 

\subsection{Non-compact Landau-Ginzburg Models}

Now we want to discuss the link between the non-compact $N=2$ superconformal field
 theories and the IR fixed point of Landau-Ginzburg models with a superpotential of the form:
\begin{eqnarray}
W_{nc} &=& \Phi^{-l}
\end{eqnarray}
where $l$ is a positive integer. This was introduced\footnote{See also e.g.
  \cite{Ghoshal:1993qt, Hanany:1994fi} for an older and related use of
  these models in the context of two-dimensional gravity.} in
\cite{Ooguri:1995wj} and we would like to understand how far we can argue for
a formal analogy with the compact case, and if possible borrow the techniques
that have been extensively used in that 
context. See also
  \cite{Hori:2000kt, Hori:2002cd}
for a more detailed analysis of the renormalization group flow with fixed asymptotics. 
The central charge of this theory with fixed asympotics 
is $c= 3+ \frac{6}{l}$. 

It is natural to assume that the field $\Phi$ cannot take the value zero, and
that moreover the point at infinity should remain a regular point. We can then
associate to the model an operator ring $\, \IC [\Phi^{-1}]$ which should
again be divided by the ideal generated by the derivative of the
superpotential. This gives rise to a ring spanned by the $l+1$ elements
$\Phi^{0}, \Phi^{-1}, \dots, \Phi^{-l}$.  Because the target-space is
non-compact, $\Phi^{0}$ is not normalizable. We exclude the operator from the
ring.  The ring of elements spanned by $\Phi^{-1}, \Phi^{-2}, \dots,
\Phi^{-l}$ is a ring without unit element. This fact contrasts with the
compact case, and it is associated to the non-existence of an $SL(2,R)$
invariant ground state in the conformal field theory. The operators
$\Phi^{-1}, \Phi^{-2}, \dots, \Phi^{-l}$ have R-charge
$\frac{1}{l},\frac{2}{l},\dots,\frac{l}{l}$.  We can match the operators
$\Phi^{-2}, \dots, \Phi^{-l}$ onto the chiral ring of the diagonal linear
dilaton theory with $N=2$ superconformal symmetry (under the assumptions of
relevance and normalizability, as discussed in section \ref{local}).  We
moreover expect the operator $\Phi^{-1}$ to become the almost-normalizable
chiral-chiral primary in the infra-red fixed point theory.

\begin{figure}
\centering
\includegraphics{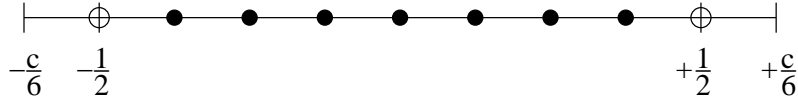}
\caption{The R-charges of the ground states in the Ramond sector
for a $N=2$ superconformal theory with central charge $c>3$.
There are almost-normalizable ground states at charges $\pm \frac{1}{2}$.}
\label{Rgroundstatesnoncomp}
\end{figure}

Again, we can summarize the chiral-chiral spectrum in a Poincar\'e polynomial
 for the $(c,c)$ ring which for the case of
(almost-)normalizable elements is:
\begin{eqnarray}
Tr_{(c,c)} t^{Q} \tilde{t}^{\tilde{Q}} &=& (t \tilde{t})^{\frac{1}{l}} + (t \tilde{t})^{\frac{2}{l}}
+ \dots +  (t \tilde{t})^{\frac{l}{l}}
\nonumber \\
& =&  (t \tilde{t})^{\frac{1}{l}}  \frac{ 1- (t \tilde{t}) }{ 1 - (t \tilde{t})^{\frac{1}{l}}}.
\end{eqnarray}
For the strictly normalizable elements, we should use:
\begin{eqnarray}
Tr_{(c,c)} t^{Q} \tilde{t}^{\tilde{Q}} &=& (t \tilde{t})^{\frac{2}{l}} + (t \tilde{t})^{\frac{3}{l}}
+ \dots +  (t \tilde{t})^{\frac{l}{l}}
\nonumber \\
&=&  (t \tilde{t})^{\frac{2}{l}}  \frac{ 1- (t \tilde{t})^{\frac{l-1}{l}} }{ 1 - (t \tilde{t})^{\frac{1}{l}}}.
\end{eqnarray}
Moreover, the normalizable Ramond ground states now have charges that go from $-1/2+1/l$ to
$+1/2-1/l$ which again lie inside the interval $]-\half, +\half[$. The almost-normalizable ground state is an outlier at charge $-1/2$. It has a spectrally flowed partner at opposite charge $+1/2$. The Ramond ground states do not reach the charge $-c/6$ and $+c/6$. See figure
\ref{Rgroundstatesnoncomp}.
The R-charges of the strictly normalizable 
Ramond-Ramond ground states can be coded in the
polynomial:
\begin{eqnarray}
Tr_{RR} t^{Q} \tilde{t}^{\tilde{Q}} &=&(t \tilde{t})^{-\frac{1}{2}+\frac{1}{l}} + (t
\tilde{t})^{-\frac{1}{2} +\frac{2}{l}}+ \dots + (t \tilde{t})^{+\frac{1}{2}-\frac{1}{l}}.
\end{eqnarray}

To summarize, we can associate a Landau-Ginzburg model to each factor of a
non-compact Gepner model. The
total superpotential will be the sum of the individual superpotentials. The
Landau-Ginzburg models we discuss will be of the Fermat type. However, we note
at this stage that there is a difference between the product
of minimal and linear dilaton theories and the corresponding product of
Landau-Ginzburg theories. Indeed, while the diagonal A-type minimal models are
identified as infra-red fixed points of Landau-Ginzburg models, the
non-compact
Landau-Ginzburg model gives rise to a deformation of the
$N=2$ linear dilaton theory.
To the linear dilaton asymptotics, we add a deforming 
potential.

So far we have studied simple Landau-Ginzburg theories (and their direct
product theories). However, to provide Landau-Ginzburg analogues of the Gepner
conformal field theories, we need to discuss orbifolded Landau-Ginzburg
models.

\section{Landau-Ginzburg Orbifolds}
\label{orbifoldedLG}

Our discussion of orbifolded Landau-Ginzburg models is mainly based on
\cite{Vafa:1989xc, Intriligator:1990ua}. The orbifolding can arise due to the
GSO projection in string theory, or due to a further geometric orbifolding of
the resulting theory. In this section, we start out by discussing the orbifold
action as being independent of possible actions on flat space factors,
following \cite{Intriligator:1990ua}. We will comment on the relation to the
full GSO projection (which also acts on the flat space factors) 
in the appendix. Since
our discussion will be closely related to the long discussion in the
literature of the compact case, we will briefly review that discussion and we
will only treat in more detail the crucial differences that exist in the
non-compact case.

For each compact Landau-Ginzburg model with superpotential $\Phi_i^{k_i}$,
there is a canonical diagonal action \be \Phi_i \rightarrow e^{ \frac{2 \pi i
  }{k_i} } \Phi_i \,, \ee which generates a $\IZ_{k_i}$ group. The exponent is
determined by the $U(1)_R$ charge of the field $\Phi_i$. For every non-compact
factor with superpotential $\Phi_j^{-l_j}$, the canonical action is \be \Phi_j
\rightarrow e^{- \frac{2 \pi i}{l_j}} \Phi_j\,.  \ee We act with the opposite
phase, since the $U(1)_R$ charge of the field $\Phi_j$ is negative (becuase of
the negative power in the superpotential). We introduce the charges $q_i$ for
the fields $\Phi_i$ for both compact and non-compact factors, 
which are equal
to \be \left(\frac{1}{k_1}, \frac{1}{k_2}, \dots ; -\frac{1}{l_1},
  -\frac{1}{l_2}, \dots\right)\,.  \ee Then the above actions on the fields
can be written as \be \Phi_i \rightarrow e^{2 \pi i q_i} \Phi_i\,.  \ee The full
symmetry group is then \be D=\prod_i \IZ_{k_i} \times \prod_j \IZ_{l_j}\,.
\ee Note that it is identical to the group we identified previously in the
context of non-compact Gepner models in section \ref{symGroup}.

We study orbifolds of the theory by a subgroup of the above diagonal group $D$.  For instance, we can choose to orbifold by a group generated by a single element, which has different weights in
each of the factors, or by a product of such groups. For an element $h$ of the orbifold group, the action on each superfield can be written as 
\be
\Phi_i \rightarrow e^{2 \pi i \Theta_i^h} \Phi_i\,.
\ee 
The phases $\Theta_i^h$ parameterize the group elements $h$.

The integer R-charge orbifold discussed in section \ref{integerR} is special,
since it is part of the GSO projection. In this case, we orbifold by the
group generated by $g_{0}$. The action of this operator on the superfields is coded as
$\Theta_i^{g_{0}}=q_i$.

Our main objective is to compute the number and charges of the
chiral/chiral $(c,c)$ and anti-chiral/chiral $(a,c)$ states, as well as their
$(a,a)$ and $(c,a)$ partners, in the orbifolded Landau-Ginzburg theory.
In particular, those with
$U(1)_R$ charges $\pm 1$ will lead to massless fields in
spacetime. In the case where the geometric approximation is valid, marginal
$(c,c)$ states correspond to complex structure deformations, and marginal
$(a,c)$ states to K\"ahler moduli.

A crucial observation is that for the Calabi-Yau examples we consider in this paper, $(c,c)$ states are in one-to-one correspondence with Ramond-Ramond ground states \cite{Lerche:1989uy}. More precisely, a Ramond-Ramond ground state becomes a $(c,c)$ state under the action of half-unit left-right symmetric spectral flow. Similarly, $(a,c)$ states are derived from Ramond-Ramond ground
states by half-unit left-right antisymmetric spectral flow. It will prove a good strategy to work in the R-R sector rather than in the NS-NS sector. The procedure we will use is the following 
\cite{Intriligator:1990ua}:
\begin{itemize}
\item[\_]First identify the unprojected Ramond-Ramond ground states in each twisted sector and compute their R-charge.
\item[\_]Then flow these states to the NS-NS sector, and keep those that are
  invariant under the orbifold action.
\end{itemize}
Let us study the two steps of this procedure in greater detail.

\subsection{R-charges of Ramond-Ramond ground states}
We first extend a result of reference \cite{Vafa:1989xc} for the twisted
sector Ramond-Ramond ground states. Consider the sector of the theory twisted
by $h$ (the following is also valid in the untwisted sector, with $h=1$).
We define a particular Ramond-Ramond ground state $|0 \rangle^h_R$, which
is the Ramond-Ramond ground state with the lowest left-moving R-charge for the
compact factors, and the highest left-moving R-charge for the non-compact
factors. The left-moving R-charges with respect to the individual factors of
that
Ramond-Ramond ground state is:
 \begin{align}\label{twistedRchargeleft}
Q &= +\sum_{\Theta_i^h \notin Z} \left( \Theta_i^h - {[}  \Theta_i^h {]} -
\frac{1}{2} \right) \cr
& \qquad+ \sum_{\Theta_i^h \in Z} (q_i - \frac{1}{2})\,,
\end{align} 
and the right-moving R-charge is:
 \begin{align}\label{twistedRchargeright}
\tilde{Q} &= -\sum_{\Theta_i^h \notin Z} \left(  \Theta_i^h - {[}  \Theta_i^h {]} -
\frac{1}{2}\right)\cr
& \qquad+ \sum_{\Theta_i^h \in Z} (q_i - \frac{1}{2})\,.
\end{align} 
Here, the expression $[\Theta]$ is defined as  the
greatest integer smaller than $\Theta$ (for $\Theta$ not an integer). Remember
that $\Theta_i^h$ is the phase that defines the action of $h$ on $\Phi_i$, and
$q_i$ is the R-charge of $\Phi_i$.

The arguments leading to the first line of these formulae are elaborate
\cite{Vafa:1989xc}. We refer to that reference for details. Note that it is
natural to look for the Ramond-Ramond ground state in the $h$-twisted sector
by twisting the left-movers half-way in one direction, and twisting the
right-movers half-way in the other direction, due to the symmetry between
left- and right-movers in the original theory. The subtraction of the integer
part of the twist arises because of the fact that we can otherwise find a
state with smaller R-charge, between $-1/2$ and $+1/2$ that will have lower
conformal dimension.  See also figures \ref{Rgroundstatescomp} and
\ref{Rgroundstatesnoncomp}. Furthermore, as argued in \cite{Vafa:1989xc}, the
R-charges behave very much like fermion number, which is indeed lifted from
$-1/2$ in the compact sector, by the contribution of $\Theta_i^h$. The crucial
difference with the compact sector is only that for the non-compact sector we
need to keep in mind that $\Theta_i^h$ is negative, and that therefore it is
more natural to think of the state with fermion number or $U(1)_R$ charge
moving {\em down} to a charge just below $+1/2$. That is precisely what is
automatically coded in the above formula, since for $\Theta$ negative but
bigger than $-1$, we have that $\Theta-[\Theta]-1/2=+1/2+\Theta$.

The second line in each of these formulae \eqref{twistedRchargeleft} and \eqref{twistedRchargeright} arises from spectral flowing the untwisted factor NS-NS ground state to the R-R sector. Note that the
superfield living in an untwisted factor can be given a nonzero vev. This is
not possible in the twisted factor, since a constant does not satisfy the
twisted boundary conditions. In this 
way, other Ramond-Ramond ground states can be generated in the same
sector. For example, if the $i$-th factor is untwisted, the state
$\Phi_i^{p}|0 \rangle_{R}^{h}$ 
($p$ integer) is another valid Ramond-Ramond ground state. It has left- and
right-moving R-charge $(Q+p|q_i|,\tilde{Q}+p|q_i|)$, where we took into
account the contribution of $\Phi_i^{p}$. In a compact factor, $p$ is
restricted to the bounds $0 \le p \le k_i-2$. In a non-compact factor,
strictly normalizable states have $2 \le -p \le l_i$, and almost-normalizable
states have $p=-1$. These bounds follow from our discussion in section \ref{LG}.

The upshot is that the formulas \ref{twistedRchargeleft} and \ref{twistedRchargeright},
derived in \cite{Vafa:1989xc} for compact models, are also valid for non-compact models.

Once this  hurdle is ovecome, one can extend the reasoning of \cite{Intriligator:1990ua} to
cover the non-compact case as well. We will not repeat the whole analysis
here. Let us observe that all the information about the unprojected
Ramond-Ramond ground states in the $h$-twisted sector is conveniently encoded
in a Poincare polynomial equal to:
\begin{align}
\Tr_{R, \text{twisted, unprojected}} t^{Q} \tilde{t}^{\tilde{Q}} & = 
\left(\frac{t}{\tilde{t}} \right)^{ \sum_{\Theta_i^h \notin Z} ( \Theta_i^h  -  [  \Theta_i^h  ] - 1/2)}
\left(t \tilde{t} \right)^{\sum_{\Theta_i^h \in Z} (q_i-1/2) }  \nonumber \\
& \times \prod_{\begin{array}{c} \Theta_i^h \in Z \\ \scriptstyle{compact}\\ \scriptstyle{factors}
    \end{array}}
\frac{1-(t\tilde{t})^{\frac{k_i-1}{k_i}}}{1-(t\tilde{t})^{\frac{1}{k_i}}}
\prod_{\begin{array}{c} \Theta_i^h \in Z \\ \scriptstyle{noncompact}\\ \scriptstyle{factors}
    \end{array}}
(t\tilde{t})^{\frac{2}{l_i}}\frac{1-(t\tilde{t})^{\frac{l_i-1}{l_i}}}{1-(t\tilde{t})^{\frac{1}{l_i}}}
.
\end{align}

In the untwisted sector, the above analysis simplifies. The particular
Ramond-Ramond ground state we discussed earlier had charges $(-c/6,-c/6)$. It
is the state one gets by half-unit spectral flow of the NS-NS vacuum. All the
superfields are untwisted, and can be given 
constant values. The relevant Poincar\'e
polynomial is simply the product of the Poincar\'e polynomials of each
factors.

\subsection{Spectral flow to the NS-NS sector}
In this way we can enumerate all the Ramond-Ramond ground states,
and their R-charges. Then we can generate all the NS-NS (anti-)chiral states,
by spectral flow.

We obtain $(c,c)$ states by symmetric half-unit spectral flow. Their R-charges are easily computed from those of the Ramond-Ramond ground state: we add $(\frac{c}{6},\frac{c}{6})$ to the R-charges of the ground state. This operation does not change the twist: $h$-twisted Ramond-Ramond ground states flow to $h$-twisted $(c,c)$ states.

In a similar way, we get $(a,c)$ states by anti-symmetric half-unit spectral
flow. This time, the Ramond-Ramond ground state R-charges are shifted by 
$(-\frac{c}{6},\frac{c}{6})$. Note that anti-symmetric spectral flow changes the
twist \cite{Vafa:1989xc} such that $h$-twisted Ramond-Ramond ground states flow to 
$h g_{0}$-twisted $(a,c)$ states.

Among all these states, we keep only those that are invariant under the orbifold
action. In the case of the integer R-charge orbifold, the selection is easy:
 we keep the states that have integer R-charges. But in other cases, 
the procedure is harder. We refer to \cite{Intriligator:1990ua} for a generic discussion of
 this projection.

\subsection{Normalizability of non-compact twisted states}

For the twisted Ramond-Ramond ground states, we encounter a subtle point and
this is the crucial difference between the compact and the non-compact case.
When we twist a coordinate, and restrict to constant modes, the field is set
to zero. This will not give rise to a normalizable state when the
Landau-Ginzburg potential has a negative power, since the potential blows up
at zero.  Although we could therefore immediately discard the states that have
a non-compact twisted factor as not being contained in the normalizable
spectrum, we advocate keeping an open mind -- we will continue the
investigation into these non-normalizable states, in view of the possibility
that we can tune the coefficient of the negative power potential to zero,
which will render these states normalizable.

Strictly speaking, the models where we do take into account such twisted
sector deformations should be thought of as existing in the well-defined
linear dilaton conformal field theories (which give rise to locally strongly
coupled string theory backgrounds). We believe we provide ample justification
for this procedure later on. In particular in section \ref{CFTmirror} we will
find that these states are needed in the construction of mirror pairs of
conformal field theories, by generalizing the Greene-Plesser prescription of
finding mirrors by orbifolding. More precisely, states untwisted in the
non-compact directions will be mapped in the mirror model to states twisted in
the non-compact directions.

To summarize, the full set of chiral and anti-chiral ring elements of the
Landau-Ginzburg orbifold, including those arising in both the untwisted and
twisted sectors, is interpreted as the spectrum of the linear dilaton theory.
The Landau-Ginzburg model is a complex structure deformation of this linear
dilaton background. This deformation removes from the spectrum the RR states
twisted in the non-compact directions and the de-singularized string
background only contains purely untwisted states. As we will see later on, in
the mirror model, the deformation is mapped to a resolution of the theory. The
spectrum of the resolved theory will, instead, only contain the states twisted
in the non-compact directions.

\section{Concrete Models}
\label{examples}
In this section we will apply what we have learned in the previous sections to
some concrete models. We  supplement the analysis of the
asymptotic partition function of section \ref{asymptotic}
with an analysis of the localized states (which we briefly touched upon 
in section \ref{local}). We start out with some
observations on the nature of non-compact Gepner models, and how they differ
from their compact counterparts. 
We will then compute the allowed deformations of $(c,c)$ and $(a,c)$ type in
particular examples, in the language of orbifolded Landau-Ginzburg models. In the conformal field theory language, the conditions to be satisfied for chiral primaries have been written out in a series of papers \cite{Eguchi:2003ik, Eguchi:2004yi, Eguchi:2004ik}. We compare our results with the conformal field theory formalism in appendix \ref{CFTanalysis}.

\subsection{Gepner Points At Large Level}

We start out with some comments that allow us to single out some particularly
interesting models.
Compact Gepner models have factors with a central charge which is always
smaller than three. A large level limit for all factors
 necessarily increases the central charge of the Gepner model, and
therefore is not consistent with the criticality condition for string
theory. Thus, compact Gepner models are necessarily at large curvature (and small
volume).

Non-compact Gepner models are of a different type. In particular we can have
non-compact Gepner models that contain factor conformal field theories at
central charge smaller {\it and} larger than three. We can therefore cancel
off the difference in a large level limit, if we wish.  Assuming that we
demand the existence of a limit in which all levels are large, we immediately
conclude that such a non-compact Gepner model necessarily has three
factors (since the total central charge is $c=9=3\times 3$).

There are two such classes of models (if we do not admit models with central
charge
precisely equal to three). One is where we have two minimal
model factors and one non-compact factor, and the other has one
minimal model factor and two non-compact factors. There are no other
possibilities at central charge $c=9$. (When we consider a non-compact
Gepner model at central charge $c=3.D$, with $D$ an integer different from
$D=3$,
there are other possibilities which can also easily be classified.)

The class of models with two minimal models can be parameterized by the 
integer levels $k_1$ and $k_2$ of the minimal models, up to
an initial choice of ADE modular invariant and a possible orbifold by a
symmetry group.
The level $l$ of the non-compact factor is then fixed to be $l
= \frac{k_1 k_2}{k_1 + k_2}$ which can be integer or fractional.

The other class of models is parameterized by two levels $l_1$ and
$l_2$ for the non-compact models, and the combination $k= \frac{l_1
l_2}{l_1+l_2}$ then needs to be integer, in order for the compact
model at level $k$ to exist. The levels $l_i$ can a priori be
fractional or real. As we have indicated before, we
concentrate on the case where all levels are integer.

\subsection*{Remarks}
\begin{itemize}
\item We note that a general class of models can be found by demanding
  $\frac{k_1 k_2}{k_1 + k_2}$ to be integer (with $k_{1}$ and $k_2$ positive
integers). Suppose we isolate the 
greatest common divisor $d$  of $k_1=d \tilde{k}_1$ and $k_2= d \tilde{k}_2$
(with $\tilde{k}_1$ and $\tilde{k}_2$ mutually prime). Then it can be shown
that the level $ \frac{k_1 k_2}{k_1 + k_2}$ is integer if and only if $d$ is a
  divisor of $\tilde{k}_1+ \tilde{k}_2$. That easily generates a large class of
  models of which we will only study a few.
\item We note that although the local curvature would seem to become small in
  the large level limit, this reasoning does not take into account the GSO
  orbifold that still needs to be performed. In specific examples it can be
  checked that the GSO orbifold recreates small radii in the geometry (see
  e.g. \cite{Israel:2004ir}\cite{Israel:2004jt} for a detailed discussion).
  We can therefore generically expect the large level limit to correspond to
  an {\em orbifolded}  weakly curved background.
\end{itemize}

In the following we will mainly concentrate on the set of models that have
the special property of allowing for a large level limit (although our
formalism does apply more widely).
We study in detail the set of models with three factors and levels $(2k,2k;k)$
or $(k;2k,2k)$ for the minimal model and non-compact factors respectively, and
orbifolds thereof. 
As a warm-up exercise however, we treat the instructive
example at complex dimension $D=2$ with the compact model at level $k$ and the
non-compact model at the same 
level $k$.

\subsection{The $(k; k)$ Model}

Let us apply the formalism for orbifolded Landau-Ginzburg models of section
\ref{orbifoldedLG} to the case of two factors, one compact and one non-compact
at equal levels $k$. We refer to the model as the $(k;k)$ model. 

The Landau-Ginzburg model has a potential 
\be
W_{LG}=\Phi_1^k + \Phi_2^{-k}
\ee
for two chiral superfields $\Phi_{1}$ and $\Phi_2$. The orbifold group that is
necessary to implement GSO is generated by 
\be
g_{0}: (\Phi_1,\Phi_2) \rightarrow (e^{2 \pi i /k} \Phi_1,e^{- 2 \pi i /k} \Phi_2)\,. 
\ee 
We will be mostly interested in obtaining the numbers and types of
deformations of the conformal field theory. In order to do this, one computes the left and right
R-charges of the Ramond-Ramond ground states in all the twisted sectors using
equations \eqref{twistedRchargeleft} and \eqref{twistedRchargeright}. Then, after
spectral flow, one can compute the R-charges of the operators in the $(c,c)$
and $(a,c)$ rings. For this $c=6$ theory, half-unit spectral flow amounts to adding $\pm
1$ unit of R-charge to the RR states.

We recall that for the $(a,c)$ states, asymmetric spectral flow from the RR
sector adds a twist by $g_{0}$.
We tabulate the R-charges of the relevant states below, and indicate with a
star the sectors in which 
non-zero constant modes can be given to the fields. We label by $\alpha$ the sector
twisted by $g_{0}^{\alpha}$.
\be
\begin{array}{c|c|c|c}
\alpha & RR & (c,c) & (a,c) \cr
\hline
0 & (-1,-1)* & (0,0)* & (-1,+1) \cr
1 & (0,0) & (+1,+1) & (-2,0)* \cr
2 \le \alpha \le k-1 & (0,0) & (+1,+1)& (-1,+1) \cr
\end{array}
\ee
We summarize the results:
\begin{itemize}
\item In the untwisted sector, the fields can have non-zero
  constant modes. The Ramond-Ramond ground state flows in the $(c,c)$ ring to the
  identity operator, with charges $(0,0)$. We moreover find $k-1$ marginal
  $(c,c)$ states in this sector. These
  arise from the monomials $\Phi_1^n \Phi_2^{-k+n}$ for $n=0,1, \dots, k-2$.
  They can be checked to be invariant under the orbifold projection.
\item Asymmetric spectral flow from the RR untwisted sector gives
  $g_{0}$-twisted $(a,c)$ states. Therefore we get $k-1$ marginal $(a,c)$
  states in this sector. The fact that these states are marginal is a
  consequence of the special value of the central charge, $c=6$.
\item For any value of the twist $\alpha=1,\dots,k-1$, we have no untwisted
  fields
in the RR sector. 
Namely the fields $\Phi_1$ and $\Phi_2$ always have twisted boundary
conditions in any twisted sector. Using the formula (\ref{twistedRchargeleft}), we find that the R-charges
  of the $\alpha$-twisted sector Ramond ground states is $(\alpha/k -0 -1/2) +
  (-\alpha/k+1-1/2) = 0$ on the left and $0$ on the right. We already see a
  phenomenon typical to our non-compact Gepner models. The twist contribution
  to the Ramond sector charges can cancel between compact and non-compact
  factors. We are in a special case, in which the twisted R-charges of all
  ground states  are zero.
\item After flowing symmetrically to the NS-NS sector, we find a single $(c,c)$ state with charges
$(+1,+1)$ in each twisted sector. They give $k-1$ marginal $(c,c)$ states in total in the twisted
sectors.
\item  Asymmetric spectral flow of the twisted RR ground states leads to $k-1$
   $(a,c)$ deformations, one in each sector (except for $\alpha=1$).
\end{itemize}

Therefore we have a total of $k-1$ marginal $(c,c)$ and $k-1$ marginal $(a,c)$
 states from the spectral flow of the untwisted RR sector ground states. In
 the Landau-Ginzburg model with potential $W_{LG}=\Phi_1^k+\Phi_2^{-k}$ these
 are the only admissible localized modes. The potential term $\Phi_2^{-k}$
 makes sure that the untwisted polynomials are allowed in the sense that they
 are normalizable at weak coupling, and have a mild behaviour at the $\Phi_2
 \approx 0$ end compared to the potential. These give rise to a $4(k-1)$
 real-dimensional moduli space of backgrounds in string theory with sixteen
 supercharges. 
In contrast, the twisted operators are not normalizable in the
 Liouville deformed model.

\subsection{The $(2k,2k;k)$ Model.}\label{twoMM}

For this slightly more complicated example, we list the full set of
(unprojected)
$(c,c)$
and $(a,c)$ states and their charges, and then pick out those that are marginal (and invariant
with respect to the orbifold projection). Again, we  perform this
exercise in the formalism of section \ref{orbifoldedLG} for orbifolded
Landau-Ginzburg models.
We consider a Landau-Ginzburg model with fields $\Phi_{1,2,3}$ and
superpotential 
\be
W_{LG} = \Phi_1^{2k} + \Phi_2^{2k} + \Phi_3^{-k}\,.
\ee 
We perform the integer R-charge orbifold, generated by the $g_{0}$:
\be
g_{0}: (\Phi_1,\Phi_2,\Phi_3) \rightarrow (e^{2 \pi i /2k} \Phi_1, e^{2 \pi i
  /2k} \Phi_2, e^{- 2 \pi i/k} \Phi_3) \,. 
\ee
In order to consider all the marginal operators, we will write down the
relevant Poincare polynomials in each sector. First of all, in the untwisted
sector we have the (unprojected, strictly
normalizable) Poincare polynomial: 
\begin{eqnarray}
\left( \frac{1-(t \tilde{t})^{(2k-1)/2k}}{1-(t
    \tilde{t})^{\frac{1}{2k}}}\right)^2  (t \tilde{t})^{\frac{2}{k}}
\frac{1-(t \tilde{t})^{\frac{k-1}{k}}}{1-(t \tilde{t})^{\frac{1}{k}}}
\end{eqnarray}
which contains $(c,c)$ states only.
When we label the twisted sectors by $\alpha=1,2, \dots, 2k-1$, we find that
for $\alpha$ smaller than $k$ there are further twisted $(c,c)$ states in the
NSNS 
sector determined by the polynomials:
\be
\left( \frac{t}{\tilde{t}} \right)^{-1/2} (t \tilde{t})^{3/2}
\ee
and when $\alpha$ is larger than $k$ by the polynomial
\be
\left( \frac{t}{\tilde{t}} \right)^{+1/2} (t \tilde{t})^{3/2},
\ee
while at $\alpha=k$ we find the polynomial
\be
(t\tilde{t})^{-1/k-1/2}  (t \tilde{t})^{+3/2} (t \tilde{t})^{\frac{2}{k}}
 \frac{(1-(t \tilde{t}))^{\frac{k-1}{k}}}{(1-(t \tilde{t}))^{\frac{1}{k}}}.
\ee
where the last factor is due to the fact that $\Phi_3$ is untwisted in this sector.
The twisted $(a,c)$ states are determined by the polynomials:
\be
\left( \frac{t}{\tilde{t}} \right)^{-2} (t \tilde{t})^0 .1
\ee
when $\alpha$ is smaller than $k$  and
\be
\left( \frac{t}{\tilde{t}} \right)^{-1} (t \tilde{t})^0 .1
\ee
when $\alpha$ is larger than $k$.
When $\alpha=k$, we get 
\be
(t\tilde{t})^{-1/k-1/2}  \left(\frac{t}{ \tilde{t}}\right)^{-3/2} (t \tilde{t})^{\frac{2}{k}}
 \frac{(1-(t \tilde{t}))^{\frac{k-1}{k}}}{(1-(t \tilde{t}))^{\frac{1}{k}}}.
\ee
One can straightforwardly determine amongst these the states that are
invariant under the orbifold action: they have integer R-charges.

\subsection*{Marginal Deformations}

We now want to look for exactly marginal deformations in these rings. These
need to have left- and right R-charge equal to $\pm 1$. As for the $(k;k)$
example, let us tabulate the R-charges of the ground states and their images
under spectral flow. Once again, $\alpha$ labels the twist. A star indicates
that some fields are untwisted and can be given nonzero vev's.
\be
\begin{array}{c|c|c|c}
 \alpha& RR & (c,c) & (a,c) \cr
 \hline
 0 & (-3/2,-3/2)* & (0,0)* & (-1,1) \cr
 1 & (-1/2,1/2) & (1,2) & (-3,0)* \cr
 2 \le \alpha <k & (-1/2,1/2) & (1,2) & (-2,2) \cr
 k & (-1/k-1/2,-1/k-1/2)* &(-1/k+1,-1/k+1)* & (-2,2)\cr 
 k+1 & (1/2,-1/2) &(2,1) & (-1/k-2,-1/k+1)*\cr
 \alpha>k+1 & (1/2,-1/2) & (2,1) & (-1,1) 
 \end{array}
\ee
The table agrees with the Poincar\'e polynomials listed previously. We find
the following marginal deformations:

\begin{itemize}
\item The untwisted $(c,c)$ marginal states are straightforwardly
  enumerated. They are given by invariant combinations of the $\Phi_i$ acting
  on the vacuum: $\Phi_1^a \Phi_2^b \Phi_3^{-c} |0 \rangle_{NS}$,
  with the bounds $0 \le a,b \le 2k-2$ and $2 \le c \le k$. The marginality
  condition is:  $a+b+2c=2k$.\\
Let's count these states. For a given $c$, each $a$ in the range $0 \le a \le
  2k-2c$ gives exactly one solution. So the total number of states is:
\be \sharp (c,c) = \sum_{c=2}^{k}(2k-2c+1) = (k-1)^2 \ee

\item A further search for marginal $(c,c)$ states gives a negative result.
  When $\alpha$ is smaller than $k$, we have $(c,c)$ states that survive
  projection, but they are not marginal since they have charges $(1,2)$. When
  $\alpha=k$, the charges left and right can also never be both equal to one. When
  $\alpha$ is larger than $k$, the charges are $(2,1)$ which also never leads to
  marginality.
  
\item Let us look for marginal $(a,c)$ states in the twisted sector. We get
  charges $(-2,2)$ when $\alpha-1$ is smaller than $k$, and therefore no marginal
  states in these sectors. When $\alpha-1$ is larger
  than $k$, we get charges $(-1,+1)$ which are marginal. So we get $(k-1)$
  $(a,c)$ states from the twisted sectors, labeled by $\{k+2,\ldots 2k\}$.
  There are no other marginal states in this theory.

\end{itemize}
In summary, we find $(k-1)^2$ untwisted marginal $(c,c)$ states, and we find
$k-1$ marginal $(a,c)$ states. Once again, only those states that arise from
the spectral flow of RR ground states with untwisted non-compact factors are
retained in the theory deformed with the Liouville potential. So all
the $(c,c)$ states are in the spectrum of the deformed theory, but none of the
$(a,c)$
 states are.

\subsubsection{Orbifolds}\label{twoMMorbifold}

We will now consider orbifolds of the above model.  As we will discuss in
detail in the next section, this exercise is useful since it generates an
infinite number of mirror theories. The logic will be analogous to the
Greene-Plesser analysis for the Gepner point in the quintic Calabi-Yau.  Under
the hypothesis that the level $k$ is not prime, we can write the level as a
product $k=k_1.k_2$ for two positive integers $k_1$ and $k_2$. Then we can
perform a $\IZ_{k_1}$ orbifold of the three-factor model that we discussed
above.

The Landau-Ginzburg model has superpotential 
$$W_{LG}= \Phi_1^{2k} + \Phi_2^{2k} + \Phi_3^{-k}.$$ 
The full symmetry group of $W_{LG}$ is $D=\IZ_{2k}
\times \IZ_{2k} \times \IZ_{k}$, where each factor acts by phase
multiplication on one of the superfields. The integer R-charge operator $g_{0}$ generates
a $\IZ_{2k}$ subgroup. It acts as: \be g_{0}: (\Phi_1,\Phi_2,\Phi_3)
\rightarrow (e^{\frac{2 i \pi}{2k}}\Phi_1,e^{\frac{2 i
    \pi}{2k}}\Phi_2,e^{-\frac{2 i \pi}{k}}\Phi_3) \,.  \ee Now consider the
$g_{1}$ operator with the following action: \be g_{1}:
(\Phi_1,\Phi_2,\Phi_3) \rightarrow (e^{+\frac{2 i \pi
    k_2}{2k}}\Phi_1,e^{-\frac{2 i \pi k_2}{2k}}\Phi_2,\Phi_3) \,.  \ee Earlier
we orbifolded the Landau-Ginzburg model by the group generated by $g_{0}$. Consider the group of order
$2k k_1$ generated by $g_{0}$ and $g_{1}$. We now orbifold the
Landau-Ginzburg theory by that group. This is compatible with
supersymmetry.
 Let us find marginal deformations in this orbifold model,
following the Landau-Ginzburg method. First
we tabulate the R-charges of the Ramond-Ramond ground states. We label the sector
twisted by $g_{0}^{\alpha} g_{1}^{\beta}$ with $\alpha$ and $\beta$, within the
bounds $0 \le \alpha \le 2k-1$ and $0 \le \beta \le k_1-1$. A star means that
some fields are untwisted.
\be
\begin{array}{c|c|c}
\alpha & \beta & RR  \cr
\hline
\alpha=0 & \beta=0 & (-3/2,-3/2)* \cr
0 < \alpha <k &  \beta k_2 < \alpha &  (-1/2,1/2) \cr
0 < \alpha <k &  \alpha < \beta k_2  & (1/2,-1/2) \cr
k<\alpha & \beta k_2 <2k-\alpha & (1/2,-1/2) \cr
k<\alpha &  2k-\alpha<\beta k_2 & (-1/2,1/2)  \cr
\alpha=0 & \beta \neq 0 & (-1/2-1/k,-1/2-1/k)* \cr
\alpha=k & \beta & (-1/2-1/k,-1/2-1/k)*  \cr
\alpha=\beta k_2 & \beta \neq 0 & (-1/2+1/2k,-1/2+1/2k)*  \cr
\alpha=2k-\beta k_2 & \beta \neq 0 & (-1/2+1/2k,-1/2+1/2k)*  \cr
\end{array}
\ee
Let us count the marginal operators:
\begin{itemize}
\item In the untwisted sector, we find $(c,c)$ chiral primaries.
  Remember that in the model orbifolded by $g_{0}$, the $(k-1)^2$ $(c,c)$ states are labeled by three
  integers $a,b,c$, such that $0 \le a,b \le 2k-2$, $2 \le c \le k$ and
  $a+b+2c=2k$. The $g_{1}$ projection keeps only those that have $a \equiv b
  \ [2k_1]$.\\
Let us work at given $c$. We want to count the number of solutions to the
  equation $a+b=2k-2c$, with  $a \equiv b\ [2k_1]$. We write $b=a+2d k_1 $,
  with $d$ integer. Then we express $a$ and $b$ in terms of $c$ and $d$ only:
  $a=k-c-d k_1$, and $b=k-c+d k_1$. The bounds on
  $a$ and $b$ imply $-k+c \le d k_1 \le k-c$. Thus we have one solution for each
  integer $d$ between $\frac{c-k}{k_1}$ and $\frac{k-c}{k_1}$.\\
So the total number of $(c,c)$ states is:
\be \sharp (c,c) = \sum_{c=2}^{k} \left( \left\lfloor \frac{k-c}{k_1}  \right\rfloor - \left\lceil
  \frac{c-k}{k_1} \right\rceil + 1 \right) \ee
where $\lfloor x \rfloor$ is the largest integer smaller or equal to $x$, and
$\lceil x \rceil$ is the smallest integer bigger or equal to $x$.\\
 This sum can be evaluated explicitly :
\be \sharp (c,c) = 
2 \left( (k_2-1)(k_1-1) + \sum_{\check{c}=2}^{k_2} (k_2-\check{c})k_1  \right) + k-1 =
k_1 k_2^2 - 2k_2 +1 \ee
\item In the sector twisted by $g_{0}^{\alpha\mp1}\ g_{1}^\beta$, we find
  $(c,a)$ and $(a,c)$ states\footnote{ The $\a\mp1$ occurs due to the shift in
    the labeling of the twisted sectors when flowing asymmetrically from the
    RR to the $(c,a)$ or $(a,c)$ sectors.}:
\begin{itemize}
\item If $0 \le \beta k_2  < \alpha <k$, or $\alpha > 2k- \beta k_2$, 
 we find one marginal $(c,a)$ state. Let's count the number of such
  twisted sectors. It can be written as:
\be \sharp(c,a) = \sum_{\alpha=1}^{k-1} \left( \left\lfloor \frac{\alpha}{k_2}
 \right\rfloor +1 \right) 
+ \sum_{\alpha=k+1}^{2k-1} \left( k_1-1- \left\lfloor \frac{2k-\alpha}{k_2}
 \right\rfloor\right) \ee
This sum can be computed explicitly as:
\be \sharp(c,a) = \left( (k_2-1)\sum_{\hat{\alpha}=1}^{k_1} \hat{\alpha} \right) +
\left( (k_1-1)(k-1) - (k_2-1)\sum_{\check{\alpha}=0}^{k_1-1} \check{\alpha}
\right) = k_2 k_1^2 -2k_1+1 \ee
\item Symmetrically, if $0 < \alpha < \beta k_2 $, or $k < \alpha < 2k-\beta k_2$, 
 we find one $(a,c)$ state. The same counting shows that there are also $k_2 k_1^2 -2k_1+1$ such sectors.
\item Eventually, if $\alpha=0$, $\alpha=k$ or $\beta k_2 = \pm \alpha \
  [2k]$, we find no marginal deformation. 
  \end{itemize}
\end{itemize}

In summary, we have in this orbifold model: 
\begin{itemize}
\item $k_1 k_2^2 -2k_2+1$ marginal $(c,c)$ states.
\item $k_2 k_1^2 -2k_1+1$ marinal $(a,c)$ states.
\end{itemize}
Once again, only the $(c,c)$ states are present in the spectrum of the
deformed theory.

\subsection{The $(k;2k,2k)$ Model.} \label{twocigars}
This example will be very similar to the previous one. For this reason our discussion will be brief and we will focus on marginal deformations. The Landau-Ginzburg model has superpotential 
\be
W_{LG}= \Phi_1^k + \Phi_2^{-2k} + \Phi_3^{-2k}\,.
\ee 
The integer R-charge operator $g_{0}$ acts on the superfields as:
\be
g_{0}: (\Phi_1,\Phi_2,\Phi_3) = (e^{\frac{2 i \pi}{k}}\Phi_1,e^{-\frac{2 i \pi}{2k}}\Phi_2,e^{-\frac{2 i \pi}{2k}}\Phi_3) \,.
\ee
We tabulate the R-charges of the Ramond ground states and their spectral
flows.
A star means that some fields are untwisted, and $\alpha$ labels the sector
twisted by $g_{0}^{\alpha}$:
\be
\begin{array}{c|c|c|c}
\alpha& RR & (c,c) & (a,c) \cr
\hline
0 & (-3/2,-3/2)* & (0,0)* & (-2,2) \cr
1 & (1/2,-1/2) & (2,1) & (-3,0)* \cr
2 \le \alpha<k & (1/2,-1/2) & (2,1) & (-1,1) \cr
k& (1/k-1/2,1/k-1/2)* &(1/k+1,1/k+1)* & (-1,1)\cr 
k+1& (-1/2,1/2) &(1,2) & (1/k-2,1/k+1)*\cr
\alpha>k+1 & (-1/2,1/2) & (1,2) & (-2,2) 
\end{array}
\ee

\begin{itemize}
\item In the untwisted sector, we find 
$(k-1)^2$ $(c,c)$ states.
\item In the sector twisted by $g_{0}^\alpha$ ($0<\alpha<2k-1$), we find one
  $(a,c)$ state if $2\le \alpha \le k$, and one $(c,a)$ state if $k \le
  \alpha\le 2k-2$.
\end{itemize}
To summarize, this model has $(k-1)^2$ marginal $(c,c)$ deformations and $k-1$
margial $(a,c)$ deformations.

\subsubsection{Orbifolds}\label{twocigarsorbifold}

We repeat the analysis of section \ref{twoMMorbifold}, and study the orbifolds
of this model. The Landau-Ginzburg model has superpotential $W_{LG}= \Phi_1^k + \Phi_2^{-2k} + \Phi_3^{-2k}$. The full group of symmetries of $W_{LG}$ is $\IZ_k \times \IZ_{2k} \times \IZ_{2k}$, where each factor acts by phase multiplication on one superfield.

The integer R-charge operator $g_{0}$ generates a $\IZ_{2k}$ subgroup and acts as follows:
\be 
g_{0}: (\Phi_1,\Phi_2,\Phi_3) \rightarrow (e^{\frac{2 i \pi}{k}}\Phi_1,e^{-\frac{2 i \pi}{2k}}\Phi_2,e^{-\frac{2 i \pi}{2k}}\Phi_3) 
\ee
Now we consider the $g_{1}$ operator with the following action:
\be 
g_{1}: (\Phi_1,\Phi_2,\Phi_3) \rightarrow (\Phi_1,e^{-\frac{2 i \pi k_2}{2k}}\Phi_2,e^{+\frac{2 i \pi k_2}{2k}}\Phi_3)
\ee
As in the $(2k,2k;k)$ model, we assume that $k=k_1. k_2$, orbifold the theory
by the subgroup generated by $g_{0}$ and $g_{1}$ and look for marginal
deformations. The 
counting is very similar 
to the $(2k,2k;k)$ case. We find
\begin{itemize}
\item $k_1 k_2^2 -2k_2+1$ marginal $(c,c)$ states.
\item $k_2 k_1^2 -2k_1+1$ marginal $(a,c)$ states.
\end{itemize}

\subsection{The $(3,3,3;2)$ Model} 

All our previous examples share an interesting feature. The marginal $(c,c)$
operators are obtained by half-unit spectral flow of untwisted RR-ground
states. On the other 
hand, the marginal $(a,c)$ operators are obtained by
half-unit asymetric spectral flow of twisted RR-ground states. A direct
consequence is that in these examples, the deformed theory only has
$(c,c)$ moduli in its spectrum. 
This statement looks general, since untwisted ground states have equal left
and right R-charges, and 
tend to flow to chiral operators. Similarly,
twisted ground states have different left and right R-charges, and would be 
expected to flow to anti-chiral operators. However, we will show in the present example
that there are exceptions to this rule.

We consider the Gepner model with three compact factors at level 3, and one
non-compact factor at level 2. The superpotential of the corresponding
Landau-Ginzburg model is \be W=\Phi_1^3+\Phi_2^3+\Phi_3^3+\Phi_4^{-2}+\Phi_5^2
\,, \ee and the model is orbifolded by the group generated by $g_{0}$: \be g_0
: (\Phi_1,\Phi_2,\Phi_3,\Phi_4,\Phi_5)\rightarrow (e^{\frac{2i\pi}{3}}\Phi_1,
e^{\frac{2i\pi}{3}}\Phi_2, e^{-\frac{2i\pi}{3}}\Phi_3, e^{-i\pi}
\Phi_4,e^{i\pi}\Phi_5) \,.  \ee The addition of $\Phi_5$ is the simplest means
to ensure that the Calabi-Yau condition is maintained while setting all the
phase factors in \cite{Intriligator:1990ua} to zero. Using the by now familiar
techniques, we tabulate the R-charges of the ground states in various sectors:
\be
\begin{array}{c|c|c|c}
\alpha & RR & (c,c) & (a,c) \\
\hline 
0 & (-3/2,-3/2)* \diamond & (0,0)* \diamond &(-1,1) \\
1 & (-1/2,1/2) & (1,2) & (-3,0)* \diamond \\
2 & (-1/2,-3/2)* & (1,0)* & (-2,2) \\
3 & (-1/2,-1/2)\diamond & (1,1)\diamond & (-2,0)* \\
4 & (-3/2,-1/2)* & (0,1)* & (-2,1) \diamond\\
5 & (1/2,-1/2) & (2,1) & (-3,1)* \\
\end{array}
\ee
The star and the diamond respectively mean that the non-compact and the
compact fields are untwisted. More precisely, $\Phi_{1,2,3}$
can have zero modes in the $\alpha=0,3$ twisted sectors for the $(c,c)$ ring
while they can have zero modes in the $\alpha=1,4$ twisted sectors in the
$(a,c)$ ring. Similarly, $\Phi_{4,5}$ can have zero modes in the $\alpha=0,2,4$
sectors in the $(c,c)$ ring, while it can have zero modes in the $\alpha=1,3,5$ sectors in the $(a,c)$ ring.

>From this, we see that there are 2 $(c,c)$ moduli:
\begin{align}
\Phi_4^{-2}\ket{0}_{c,c}^{\alpha=0} \quad \text{and} \quad \ket{0}_{c,c}^{\alpha=3}
\cr
\end{align}
and $2$ $(a,c)$ moduli:
\be
\ket{0}_{a,c}^{\alpha=0} \quad \text{and}\quad \Phi_4^{-2}\ket{0}_{a,c}^{\alpha=3} \,.
\ee
Notice that the $(c,c)$ modulus in the third twisted sector occurs in a sector
in which the fields that corresponds to the non-compact direction, $\Phi_4$,
is twisted. Moreover, the $(a,c)$ modulus in the third twisted sector appears
while the non-compact direction is untwisted. As a consequence, the theory deformed by
the Liouville potential has only one $(c,c)$ modulus in its spectrum, {\em
  and} it 
also has one $(a,c)$ modulus. 
(On the other hand, the resolved theory will have
one $(a,c)$ modulus, together with one $(c,c)$ modulus.) It can be
checked by direct calculation that this is consistent with the equations
analyzed in \cite{Eguchi:2004yi}. 
The Gepner model analysis of this model is discussed briefly in appendix \ref{3332}.

\section{Mirror Symmetry For Non-compact Gepner Models}\label{CFTmirror}

In this section we address the question of identifying mirror pairs. In the
case of compact Gepner models, when we specify the diagonal model as our
starting point, we obtain the mirror model by modding out by the maximal
discrete subgroup $H$ of the  diagonal group $G$ that is
consistent with space-time supersymmetry \cite{Greene:1990ud}. Subgroups $F$ of $H$ give rise to
models that are mirror to models modded out by $H/F$.

In the following we will argue that non-compact Gepner models behave very
similarly, in their undeformed guise. In particular, we shall show that modding
out by the maximal subgroup consistent with supersymmetry, we exchange $(c,c)$
and $(c,a)$ deformations of the undeformed theory.

A corollary of 
this statement is that a theory deformed by a given operator, will map after mirror symmetry to the mirror 
theory, deformed by the mirror operator. Thus, mirror symmetry applies to the deformed, regular theories as well. 

The main difference with the compact models is therefore that the mirror map
includes the specification of the action of the mirror map on the deforming
operator(s). Naturally, the specification is that one changes the right-moving
R-charge of the deforming operator to find the mirror deformation. The mirror map extends to subgroups of $H$ as in the compact case.

We believe it is best to illustrate the above general framework in a few
examples. In the following section, we will then revisit these examples and
see to what extent we can interpret mirror symmetry of the conformal field
theories in a geometric framework.

\subsection{Sixteen Supercharges}

\begin{figure}
\centering
\includegraphics[scale=.92]{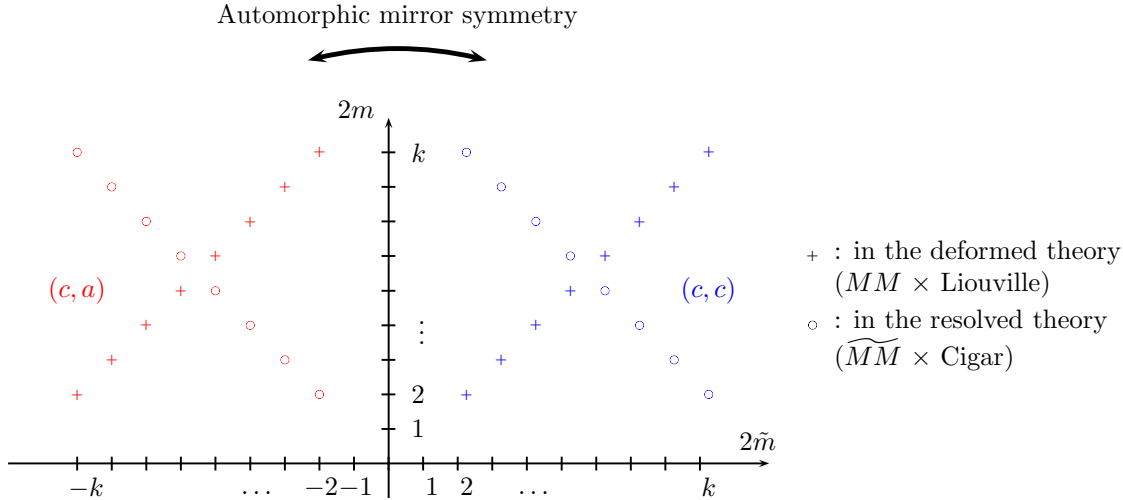}
\caption{Chiral-chiral and chiral-anti-chiral
primaries in the $(k;k)$ model. Each dot corresponds to one chiral primary,
identified by its non-compact left- and right-moving
quantum numbers $2m$ and $2\tilde{m}$.\label{ALE}}
\end{figure}

We recall that compact Gepner models at central charge $c=6$ lie in the moduli
space of $K3$ compactifications of string theory. It is well-known
\cite{Aspinwall:1994rg} that the mirror transform acts as an automorphism of
the K3 moduli space and there are special points in the moduli space where
there are fixed points. For our non-compact $(k;k)$ model at central charge
$c=6$, we see that in its singular guise, the particular $8(k-1)$ deformations
that we identified are indeed self-mirror. This can be seen in figure
\ref{ALE}. Changing the sign of the right-moving R-charge exchanges the two
individual sets of $4(k-1)$ deformation parameters that we distinguished
previously. We note that this property of self-mirroring holds only for the
singular model.\footnote{We do not advocate that the reader take the
  undeformed (well-defined) conformal field theory as a
 good description
  of the strongly coupled string background. We use it as a formal tool in
  arguing for the precise points of analogy and difference with compact Gepner
  models.}

As a consequence, we can discuss mirror symmetry for the weakly coupled deformed
model. When we deform the $(k;k)$ model with a Liouville potential (consistent
with sixteen supercharges), we are left with $4 (k-1)$ deformation parameters
and a regular theory. It is mirror dual to the singular theory deformed by the
winding potential (which geometrically gives rise to the cigar theory). The latter
theory also has $4(k-1)$ deformation parameters, which can be mapped
individually to their mirror images (using their worldsheet R-charges for each
factor of the model).

\subsection{Eight Supercharges}

For compact Gepner models at central charge $c=9$, mirror symmetry maps one model onto another, exchanging $(c,c)$ and $(c,a)$ states. We will illustrate that this is the case for non-compact Gepner models as well. First we will treat a case in which we simply mod out by the maximal group, and thus
obtain the mirror theory. Then we will show that one can mod out by a subgroup of the maximal subgroup and obtain mirror pairs. We will thus provide  large classes of mirror non-compact Gepner models.

\subsubsection{The $(k;2k,2k)$ Model}\label{zkoneorb}

We note that the $(k;2k,2k)$ model, which is modded out only by the GSO
$\IZ_{2k}$ group (in the Landau-Ginzburg formulation) has $(k-1)^2$ marginal
$(c,c)$ deformations and $k-1$ marginal $(a,c)$ deformations, as discussed in
section \ref{twocigars}. Orbifolds of this model were treated in section
\ref{twocigarsorbifold}.  The maximal group that we can divide out by
consistent with supersymmetry is the orbifold group when $k_1=k$ and $k_2=1$.
Then, substituting in the formulae for the massless moduli computed in that
section, we get $k-1$ marginal $(c,c)$ deformations and $(k-1)^2$ marginal
$(a,c)$ deformations. The orbifold indeed gives rise to the mirror theory.
This is exactly analogous to the Greene-Plesser discussion of mirror conformal
field theories associated to compact Calabi-Yau threefolds at a Gepner point.

We want to compare these models to those in the literature. The $(k;2k,2k)$
model modded out by the maximal group and deformed by the sum of Liouville
potentials corresponds to the non-compact Calabi-Yau studied in
\cite{Lerche:2000uy} to geometrically engineer pure $SU(k)$ gauge theory. The
orbifold group restricts all the possible $(c,c)$ deformations to the $k-1$
moduli studied in that paper which span the Coulomb branch of the $SU(k)$
gauge theory. We will discuss the relation between our non-compact Gepner
model, the associated Landau-Ginzburg model and this geometry in more detail
in the next section.

Moreover, we find that this model is mirror to a model that has no orbifold
except the GSO projection, and deformed by the winding condensates in the two
non-compact directions. That gives rise to the two-cigar model of
\cite{Eguchi:2004ik}, as argued in that reference. From our perspective, we
see that the double cigar deformations disallow all $(k-1)^2$ marginal $(c,c)$
deformations in the unorbifolded model, and leaves only $k-1$ K\"ahler
deformations.  That matches precisely the analysis in \cite{Eguchi:2004ik}.
Note that this gives a confirmation of our methodology: the counting in
\cite{Eguchi:2004ik} is based on the spectrum identified by considering a
regularized partition function \cite{Maldacena:2000kv, Israel:2004ir,
  Hanany:2002ev, Eguchi:2004yi}. Thus we find that the regularized partition
function agrees with our intuitive arguments which find their basis in the
Landau-Ginzburg model with negative power potentials (see sections \ref{LG}
and \ref{orbifoldedLG}).

As an example of the power of our simple description, we note that the model
with Liouville deformations and no orbifold action is not directly related to
the known models described in \cite{Lerche:2000uy, Eguchi:2004ik}.
Nevertheless, we readily identify its mirror to be the orbifolded theory with
cigar deformations in both the non-compact factors.

\subsubsection{The Orbifolded $(2k,2k;k)$ Models}

It should be clear now that the examples in sections \ref{twoMMorbifold} and
\ref{twocigarsorbifold} give rise to two infinite classes of mirror
non-compact Gepner models. We will illustrate this for one of the two classes
since the other class behaves in almost every respect analogously.\footnote{We
  will signal the exception in the next section.}

\begin{figure}
\centering
\includegraphics[scale=.92]{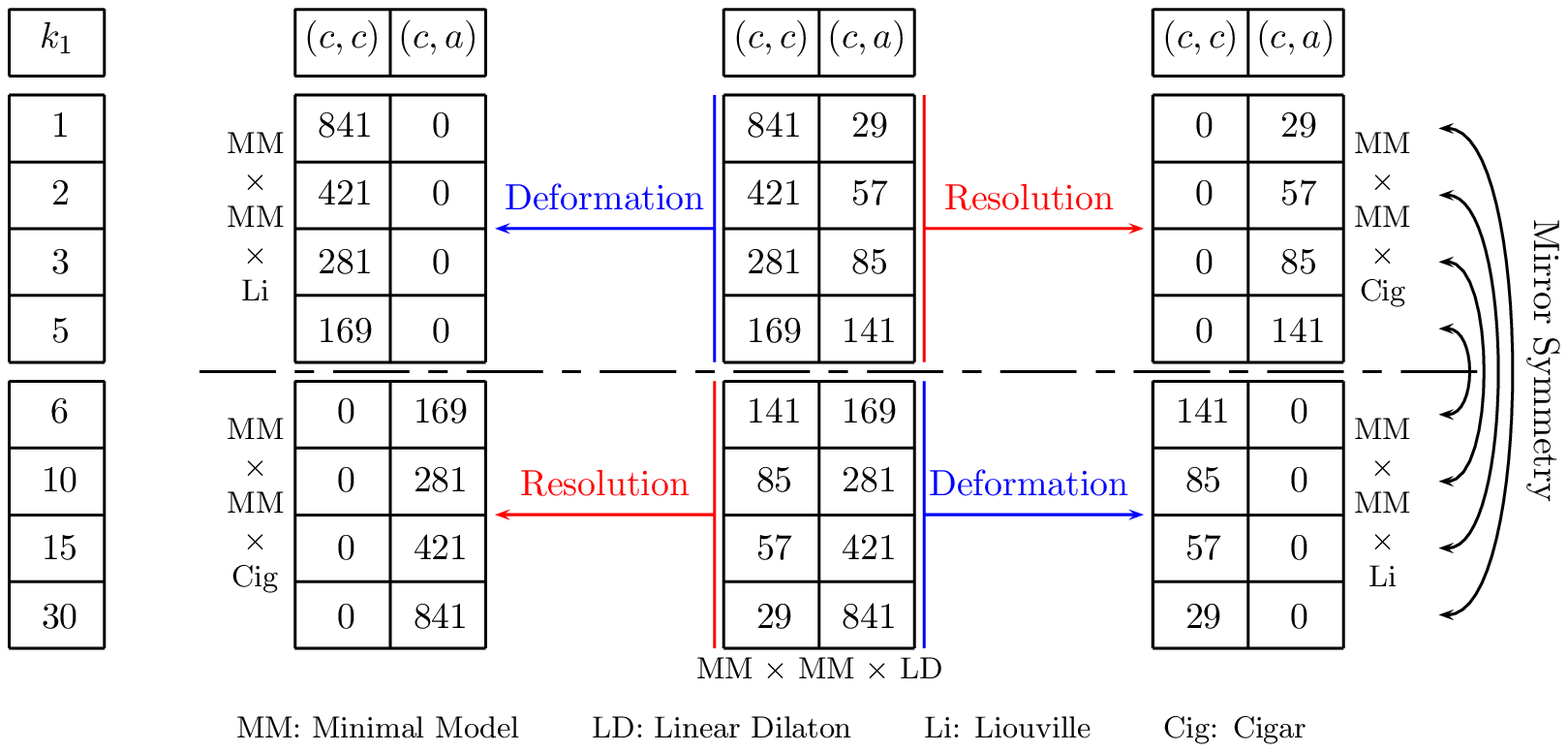}
\caption{An explicit example: the $(2k,2k;k)$ models and its orbifolds by
  $\IZ_{k_1}$, with $k=30$.}
\label{mirrorsymmetryfigure}
\end{figure}

Recall that the $(2k,2k;k)$ model allows for a maximal $\IZ_k$
orbifold
consistent with supersymmetry. When the level $k$ is a product of two positive
integers
$k=k_1k_2$,
we can orbifold by a non-maximal subgroup $F=\IZ_{k_1}$ which gives rise to a
model which is mirror to a $\IZ_{k_2}$ orbifold of the same model. Indeed, the
counting of $(k_1k_2^2-2k_2+1)$ marginal $(c,c)$ operators and
$(k_2^2k_1-2k_1+1)$ marginal $(c,a)$ states in the first model (the
$\IZ_{k_1}$ orbifold) is precisely the mirror of the counting of the model
with $\IZ_{k_2}$ orbifold group, as can be seen by a exchanging the role of
$k_1$ and $k_2$ (see figure \ref{mirrorsymmetryfigure}). Thus, we have very
simply but explicitly demonstrated the existence of an infinite number of mirror pairs of singular non-compact Gepner models.

We note that the model $(k;2k,2k)$ can be treated analogously. The proof
of the generic fact that models modded out by 
a subgroup $F$ of $H$
are mirror to models modded
out by $H/F$ runs along very much the same lines as in the compact case \cite{Greene:1990ud}, for
the undeformed theory. For the deformed theory we need to remember that the
mirror map will mirror the deformation as well.

\section{Mirror Non-compact Geometries and their Relation to $\,\IC^n/\Gamma$ Orbifolds}
\label{geometrymirror}
In this section we want to discuss some geometric realizations of the mirror
map that we identified in the non-compact Gepner models above. We will see
that we can identify various models with non-compact Calabi-Yau geometries,
and approximate their mirror duals with abelian orbifolds of $\, \IC^3$ at
large levels. At finite level, we find that the results of toric geometry 
acquire important modifications that lift certain
moduli in our models.

\subsection{The $(k;k)$ Model}

In this model, the space-time background has sixteen supercharges. In
analogy with the compact case, we identify the Calabi-Yau manifold that
corresponds to the Landau-Ginzburg model by writing down the superpotential
(augmented with the appropriate number of quadratic coordinates) in the non-compact weighted projective space $W\IC\IP^{4}$:
\be
w_1^k + w_2^2 + w_3^2 + c w_4^{-k} = 0.
\ee
The constant $c$ (multiplied by the coefficient of the first monomial)
measures the strength of the Liouville deformation.
By scaling the $\, \IC^\ast$ valued coordinate $w_4$ to one, we recuperate the
equation:
\begin{eqnarray}
z_1^k + z_2^2 + z_3^2 + c &=& 0,
\end{eqnarray}
which describes an ALE space which is deformed by $c$ from its singular
orbifold limit. We thus associate that geometry to the Landau-Ginzburg model, but we should
keep in mind that this association is local, i.e. near the singularity. 
Asymptotically the spaces differ. We will see an example of the consequences of
this difference later on.
The matching of the $4(k-1)$ marginal deformations to the geometric moduli is
well-known.

\subsubsection{The Mirror Theory}

The conformal field theory mirror to the above theory was argued to be the
$(k;k)$ model with cigar deformation. In the appendix we recall, following
\cite{Ooguri:1995wj, Kutasov:1995te}, that after a single T-duality, this
model is mapped to a configuration of NS5-branes spread on a circle in a
transverse plane. Moreover, using the explicit geometric description, it can
be argued in great detail (see the appendix) that the NS5-branes spread on the
circle map under that T-duality to an orbifold singularity of the type $\, \IC^2/\IZ_k$ at the tip
of the cigar and the center of the minimal model disc. In particular, we see
that the $\IZ_k$ orbifold that arises from the GSO projection on the side of
the cigar mirror conformal field theory acts {\em geometrically} on the cigar
and minimal model coordinates close to the tip of the cigar and the center of
the disc.

One should contrast this geometric action to the lack of such an action in the
mirror geometry. Moreover, in this example, it becomes 
manifest that when the cigar/winding deformation is turned on in the singular
theory, the deformation caps off the linear dilaton cylinder. As a consequence
it gives rise to a (geometric) fixed point which allows for the localization
of twisted sector states. That agrees with our prescription for keeping the
$4(k-1)$ twisted sector states when turning on the cigar deformation.

Note that there is a one-to-one match of the marginal supersymmetric
deformations of the cigar times minimal model conformal field theory to the
marginal supersymmetric deformations of the $\, \IC^2/\IZ_k$ orbifold (with
$B$-field on the vanishing cycles) \cite{Ooguri:1995wj}.

In the strictly infinite level limit, the match is expected, since then the
cigar and disc flatten out completely. The two models become identical in that
limit. However, at finite level $k$, the match is due to 
the rigidity of the sixteen supercharge construction, as will 
become clear when we discuss models 
with less supersymmetry.

\subsection{The $(k;2k,2k)$ Model}

Models with eight supercharges will show various new and interesting features.
We will discuss two models in detail with increasing level of complexity. The
model with levels $(k;2k,2k)$ and double Liouville deformation has a
singularity that is well-described by a hypersurface in the
(non-compact)weighted projective space of the form \cite{Eguchi:2004ik}:
\begin{eqnarray}
w_1^k + w_2^2 + w_3^2 +  w_4^{-2k} + w_5^{-2k}  &=& 0,
\end{eqnarray}
where the two coordinates $w_{4,5}$ are $\, \IC^\ast$ valued.  The deforming
monomials are in one-to-one correspondence with the $(k-1)^2$ marginal $(c,c)$
deformations that we identified previously. After performing the $\IZ_{k}$
orbifold, we find that only the complex structure
deformations of the form $w_1^n (w_4 w_5)^{-k+n}$ for $n=0,1,\dots,k-2$ are
invariant under the orbifold action. As mentioned earlier, this is precisely
the geometry discussed in \cite{Lerche:2000uy, Hori:2002cd, Eguchi:2004ik}.

\subsubsection{The Mirror Theory }

Let us now turn to the mirror conformal field theory and see whether we can count the number of K\"ahler deformations of the mirror theory using geometric means. The first observation we make is that, just as in the case with sixteen supercharges, there is an infinite level limit in which the model flattens out, up to an overall orbifold action. We refer the reader to appendix \ref{NS5} for the basic arguments in favour of such a description. In that (strict) limit, the model, locally, near the tips of the cigars and the center of
the minimal model disc becomes equivalent to the orbifold $\, \IC^3/\IZ_{2k}$. Again, we use that on this side of the mirror symmetry, the GSO projection acts geometrically and infer that the
action of the $\IZ_{2k}$ on the three factors of $\, \IC^3$ is 
weighted
as $\frac{1}{2k} (2k-1,2k-1,2)$. This orbifold is toric and the number of
K\"ahler deformations of the geometry can be counted using toric geometry
techniques\footnote{See, for instance, \cite{Aspinwall:1994ev, lustetal} and references
  therein for a review of these methods.} as follows.

Consider the supersymmetric orbifold $\,\IC^3/\Gamma$, where the orbifold group of order $N$ is generated by 
\be \theta: (z_1,z_2, z_3) \longrightarrow (\omega^{a_1} z_1,\omega^{a_2} z_2, \omega^{a_3} z_3) \quad \text{such that}\quad \sum_{i}a_i = 0 \quad \text{mod} \quad N \,.  
\ee 
The counting of K\"ahler deformations proceeds as follows.

\begin{itemize}
\item Consider all powers of $\theta$ and list their exponents in multiples of
  $2 \pi i$ such that they fall in the range $1/N \times (0,N-1)$. Label them
  $(g_1, g_2, g_3)$.
  
\item The K\"ahler moduli are in one to one correspondence with those powers
  of $\theta$ that satisfy the following conditions: \be\label{constraint}
  \sum_{i=1}^{n} g_i = 1 \quad \text{with} \quad 0\le g_i < 1 \,.  \ee The
  power of $\theta$ tells you in which twisted sector the modulus appears.
\end{itemize}

Before we proceed further let us also briefly recall how one draws the toric
diagram corresponding to the non-compact orbifold. This will turn out to be
useful to compare and contrast the spectrum of these non-compact toric
orbifolds with the Landau-Ginzburg computation of the spectrum of moduli for
the non-compact Gepner models.

Given an action of $\theta=(\theta_1, \theta_2, \theta_3)$ on $\,\IC^3$ as
above, we first find basis vectors $\{D_1, D_2, D_3\}$ that satisfy \be
\sum_{i=1}^{3}\theta_i \, (D_i)_a = 0 \quad \text{mod}\quad N \,.  \ee These
three basis vectors generate the toric fan. One solution is  $(D_i)_3 = 1\ 
\forall\ i$. One can find two other linearly independent solutions to this
equation. Thus, neglecting the third coordinate of the vectors and plotting only the other two solutions to the above equation, one gets points on a plane. That this must be so follows from the Calabi-Yau condition. These points define the Newton polyhedron corresponding to the non-compact Calabi-Yau. The basis vectors $D_i$ form a cone over the polyhedron which allows one to draw the toric diagram for the Calabi-Yau threefold in the plane. Now, for each power of $\theta^{n}=(g^{(n)}_1,g^{(n)}_2,g^{(n)}_3)$ that satisfies the above two constraints (i.e. for every K\"ahler deformation),we add another vector (interior or boundary point in the toric diagram) $E_n$ given by 
\be 
E_n = \sum_{i=1}^{3}\, g^{(n)}_i D_i \,.  
\ee

Let us apply the above algorithm to compute the K\"ahler moduli and draw the
toric diagram, for our case where $\Gamma=\IZ_{2k}$ (i.e. $N=2k$) and \be
(\theta_1, \theta_2, \theta_3) = \left(\frac{2k-1}{2k}, \frac{2k-1}{2k},
  \frac{2}{2k}\right)\,.  \ee One can check that there are $k$ K\"ahler
deformations which arise from the twisted sectors $k, k+1, \dots, 2k-1$. This
is one more than the number of $(c,c)$ deformations we obtained from the
Landau-Ginzburg computation in section \ref{twocigarsorbifold}. Let us draw
the toric diagram corresponding to this orbifold. We choose a basis of vectors
\be D_1 = (0,1,1) \quad D_2 =(0,-1,1) \quad D_3=(k,0,1) \,.  \ee 
We add $k$ points, corresponding to the $k$ K\"ahler deformations, which are elements
in the $(c,a)$ ring, in the twisted sectors, whose coordinates are given by
\be E_{k}=(0,0,1)\,, \quad E_{k+1} = (1,0,1)\,, \quad E_{k+2} = (2,0,1)\,, \quad \ldots
\quad E_{2k-1} = (k-1,0,1) \,.  \ee
\begin{figure}
\centering
\includegraphics[scale=.75]{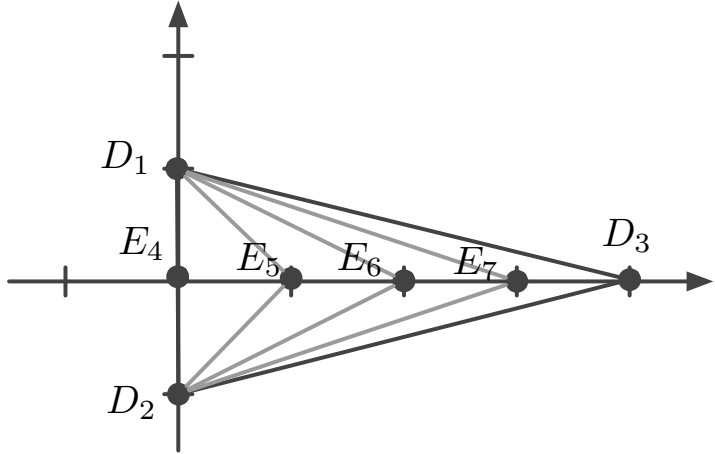}
\caption{The toric diagram for $\,\IC^3/\IZ_{2k}$ for $k=4$. Note hat $E_{k=4}$ corresponds to a boundary point while all other added points lie in the interior of the Newton polyhedron. We have also shown a possible triangulation of the polygon, corresponding to a particular resolution of the singularity.\label{toric}}
\end{figure}
These points are plotted in figure \ref{toric}. 
Since $E_k=\half(D_1+D_2)$ in our $\IZ_{2k}$ orbifold it follows that this
point will always be on the boundary of the toric diagram. 

When we compare the
two calculations of the moduli, we see that the difference arises in 
the $2k-1$st twisted sector of the GSO orbifold, since, in the conformal field
theory, we did not find a marginal K\"ahler deformation in this twisted sector.
Let us study in somewhat more detail how this difference comes about.

It is well known that the points in the toric diagram can also be associated
to exceptional divisors of the resolution of the orbifold. This is the twist-field-divisor map discussed in \cite{Aspinwall:1994ev}. The topology and
intersection numbers of these divisors can be obtained in a straightforward
manner by using the dual toric diagram. The boundary point corresponds to
non-compact divisors and have the topology $\IP^1 \times\, \IC$. Therefore,
 the existence of the resolution corresponding to 
$E_k$ in the conformal field theory might seem
surprising, given that the (strictly) normalizable deformations in the conformal field theory can be
associated to compact cycles in the geometry. This can be understood as
follows: the $k$th power of the orbifold action is trivial on one of the three
complex directions in $\, \, \IC^3$, and creates a singularity that stretches along
a complex line in $\, \, \IC^3$. The deformation is therefore akin to a K\"ahler
deformation of a $\, \, \IC^2/\Gamma$ orbifold, embedded in $\, \IC^3$. However, it is
important to note that the $\,\IC$ direction that is left invariant in the
$k$th twisted sector, is the direction that is compactified 
in the conformal
field theory by the addition of the Landau-Ginzburg
potential.
 So in the
conformal field theory, all the exceptional divisors become compact. A second
effect of the compactification is that only $k-1$ of 
the deformations are
linearly independent. The Landau-Ginzburg counting of chiral primaries gives an easy method
to understand which of the exceptional divisors are chosen as a basis in the
conformal field theory.

\begin{figure}
\centering
\includegraphics[scale=0.75]{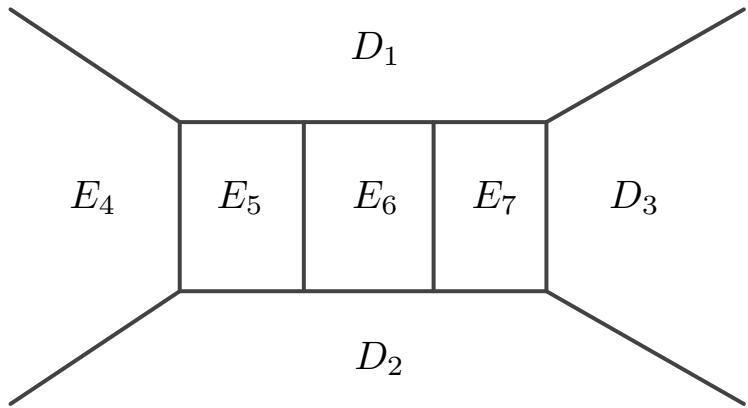}
\caption{The dual toric diagram for $\,\IC^3/\IZ_{2k}$ with $k=4$. We use the same alphabet to denote the point in the toric diagram and the divisor corresponding to it in the dual toric diagram. Note that $E_{k}$ (which is $E_4$ in our case) is non-compact while all the other exceptional divisors are compact. \label{dual}}
\end{figure}

Note that the $(c,a)$ deformation which is the identity operator in the
minimal model, sets the volume of the compact factor and simultaneously
the volume of the two-cycle at the $\, \IC^2/\IZ_{2}$ singularity.  The operator has
charges $r=(0,0,0;0;k,k)$ and $\tilde{r} = (0,0,0;0;-k,-k)$ as can be seen
from the Landau-Ginzburg model description, or appendix \ref{k2k2k}.

\subsubsection{The $(2k,2k;k)$ Model}

The toric abelian orbifold that is related to the $(2k,2k;k)$ model is the
same as the one we discussed before. The differences with the previous model
lie in the fact that we now compactify two coordinates, which reduces the
number of K\"ahler moduli by two. By analyzing the spectrum of the conformal field theory as
before, we find that in the toric diagram corresponding to the flat space
orbifold, the twist fields associated to the divisors $E_{k}$ and $E_{k+1}$ are
excluded in the conformal field theory.

However, we gain a K\"ahler modulus in the $0$th twisted sector which sets the
overall volume of the two compact factors. It is the identity operator in the
minimal model factors, and the winding operator in the cigar factor.  It has
charges $r=(0,0,0;0,0;k)$ and $\tilde{r}=(0,0,0;0,0;-k)$ as can be seen from
the Landau-Ginzburg description or from appendix \ref{2k2kk}.  This is
consistent with the picture developed in \cite{Giveon:1999px}: the effective
string coupling at the tip of the cigar sets the volume of the resolved
cycles. 

Note that in the previous example, we had a slight refinement of the picture
developed in \cite{Giveon:1999px}. Namely, in the presence of the two
non-compact directions, it is the modulus associated to the $\IZ_2$ orbifold
singularity at the tips of the cigars that sets the volume of the internal compact space.

\subsubsection{The $\IZ_{k_1}$ Orbifold Of The $(k;2k,2k)$ Model}
In order to understand some more general features of the spectra of the
conformal field theory and how they fit into the spectrum of the flat space
orbifold approximation, let us study the orbifold models studied in sections
\ref{twoMMorbifold} and \ref{twocigarsorbifold}.

\begin{figure}
\centering
\includegraphics[scale=0.75]{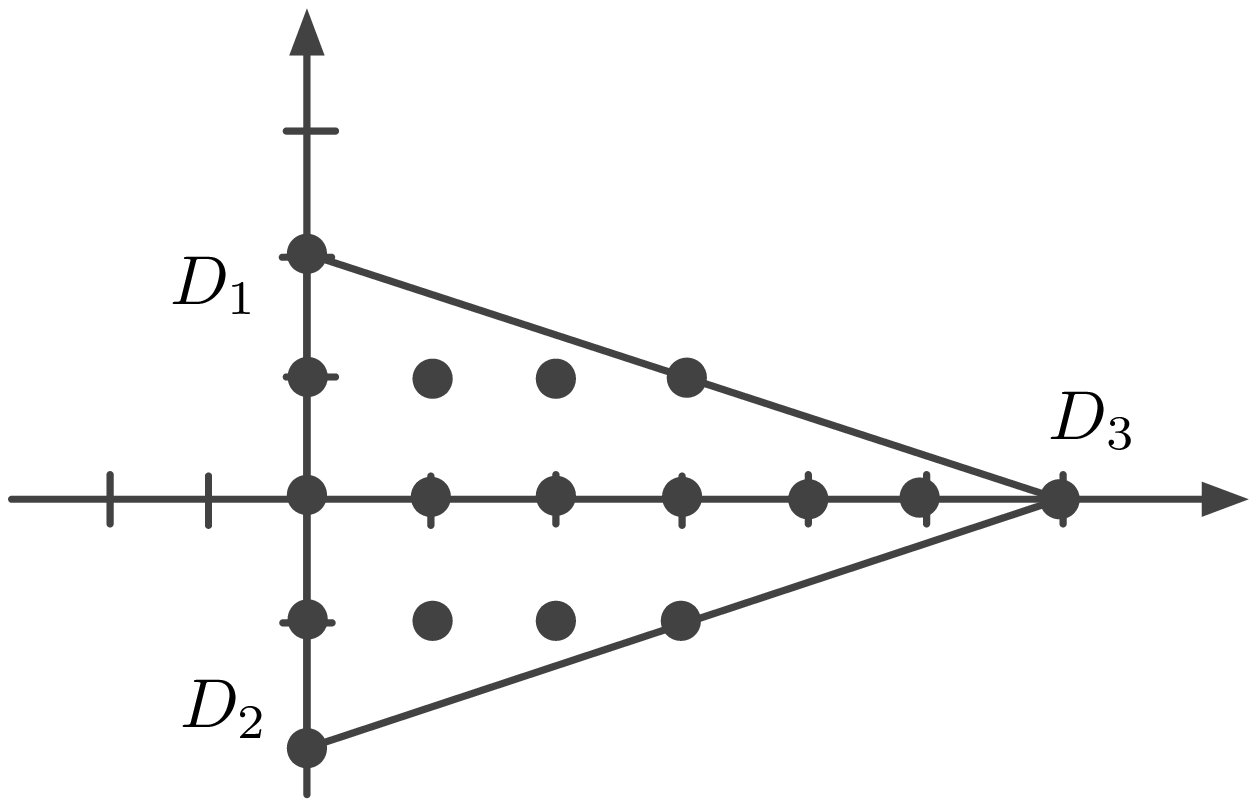}
\caption{The toric diagram for $\,\IC^3/(\IZ_{2k}\times \IZ_{k_1})$ with $k=6$ and $k_1=2$. \label{orbtoric}}
\end{figure}

The flat space approximation to the conformal field theory is given by a $\,\IC^3/\Gamma$ orbifold, generated by the elements
\begin{align}
g_0 &= \frac{1}{2k}(2k-1,2k-1,2) \quad\text{and}\cr
g_1 &= \frac{1}{2k}(k_2,2k-k_2,0) \,.
\end{align}
Using the algorithm discussed earlier, one can easily draw the toric diagram
associated to this singularity as in figure \ref{orbtoric}.
\begin{figure}
\centering
\includegraphics[scale=0.75]{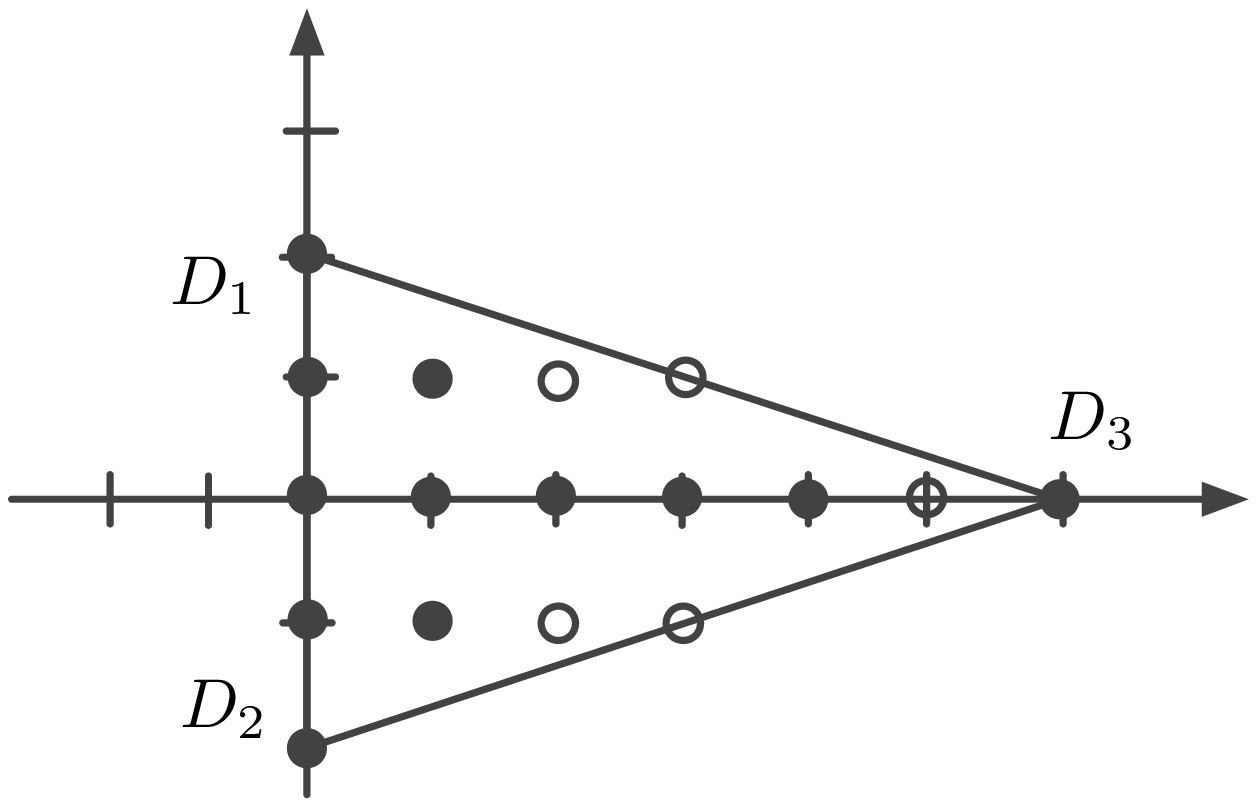}
\caption{Spectrum of the $\IZ_{k_1}$ orbifold of the $(k;2k,2k)$ conformal field theory for $k=6$ and $k_1=2$ are denoted by the filled dots. The unfilled dots show those points in the flat space approximation which are not included in the conformal field theory. \label{orbthree}}
\end{figure}
\begin{figure}
\centering
\includegraphics[scale=0.75]{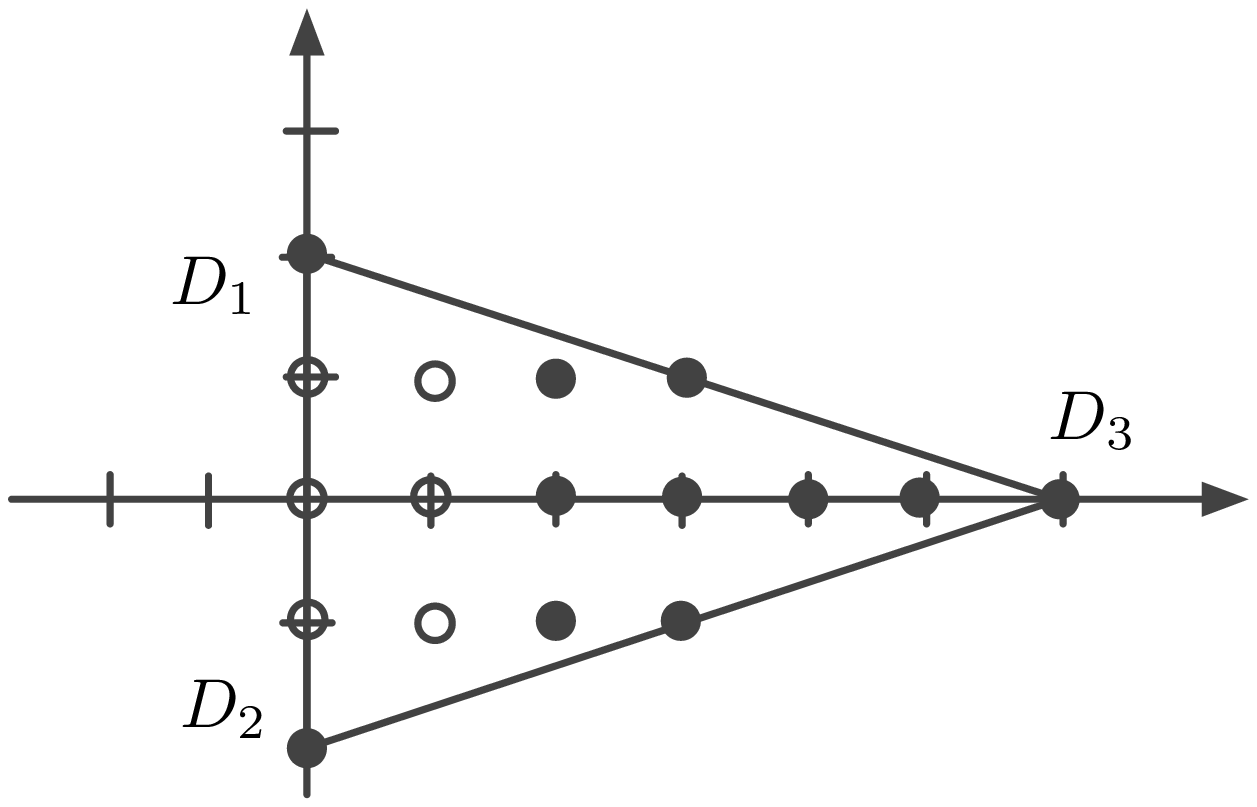}
\caption{Spectrum of the $\IZ_{k_1}$ orbifold of the $(2k,2k;k)$ conformal field theory for $k=6$ and $k_1=2$ are denoted by the filled dots. The unfilled dots show those points in the flat space approximation which are not included in the conformal field theory. \label{orbtwo}}
\end{figure}
The spectrum of the exact conformal field theory has already been discussed in the main part of
the paper. If we now plot the spectrum of the exact conformal field theory that corresponds to the $(k;2k,2k)$ model along the same lines, we get figure \ref{orbthree}.  We have shown which elements of the flat space approximation get lifted in going to the exact conformal field theory description.

\subsubsection{The $\IZ_{k_1}$ Orbifold Of The $(2k,2k;k)$ Model}
For this model the flat space approximation is identical to the one discussed
above\footnote{For the $(2k,2k;k)$ example, the canonical group element $g_0$
  is given by the inverse of the one we have written here but the group
  generated and the orbifold of $\,\IC^3$ associated to it is unchanged.  }
and it is drawn in figure \ref{orbtoric}. The conformal field theory analysis
is, however, different and the K\"ahler deformations which are kept are shown
in the figure \ref{orbtwo}.

We mentioned before that in the toric diagram for the flat space orbifolds,
the boundary points correspond to non-compact divisors. Nevertheless as we saw
in the simpler orbifold example, in the exact conformal field theory
description, some of these directions are compactified. These directions are
those along which a potential of the form $\Phi^n$, $n > 0$ is turned on and
which flow into a minimal model in the IR. However, the twist fields that correspond to divisors (via the twist-field-divisor map) which extend along directions that remain non-compact should be excluded in the conformal field theory as these do not lead to normalizable deformations. That this is
so can be checked in our examples.

For instance, in the $(2k,2k;k)$ model, in figure \ref{orbtwo}, the excluded
points on the boundary of the toric diagram correspond to K\"ahler
deformations which are in the twisted sectors $g_1^{\b}$ and
$g_0^{k}g_1^{\b}$, with $\b = 1, \ldots k_1-1$. Their respective $U(1)$
charges are given by \be \frac{1}{2k}(\b k_2,2k-\b k_2,0)\quad\text{and}\quad
\frac{1}{2k}(k+\b k_2,k-\b k_2,0)\quad \text{for}\quad n\in \{1, \ldots, k_1-1
\}\,.  \ee As one can see, the twist fields are uncharged under the $U(1)$
that acts on the non-compact direction and the divisors that correspond to
these fields are non-compact. These are subsequently excluded from the
conformal field theory spectrum.

However, further work is required to understand in full generality which states of the flat space orbifold are retained in a given conformal field theory of the type studied in this article.

\subsection{Relation to NS5-brane set-ups}
The relation of the above orbifold approximations to NS5-brane set-ups has been discussed in the literature. These toric abelian orbifolds of $\, \, \IC^3$ can  be
mapped one-to-one to NS5-brane configurations that wrap holomorphic curves. The
holomorphic curves can in turn be described by dimers that are
systematically reconstructed from the abelian orbifold group. In the
particular case above, one obtains hexagon tilings of the plane with 
labelings determined by the weights of the GSO projection and further
orbifolding. They can be determined by straightforwardly generalizing  
the examples in \cite{Hanany:2005ve, Franco:2005rj}.

In the sixteen supercharge case, one can show explicitly that the non-compact
Gepner model captures the near-horizon doubly scaled limit of the backreacted
NS5-brane geometry (as we have recalled in the appendix).
 In the case of eight supercharges, our non-compact Gepner
models capture a near-horizon doubly scaled limit of the backreaction of
NS5-branes wrapped on the holomorphic curves coded by the dimer corresponding
to our non-compact Gepner model.

\section{Conclusions and future directions}\label{conclusions}

We have shown that the Gepner formalism for constructing modular invariant
partition functions carries over to the asymptotic partition function of
non-compact Gepner models. The analogy was then further developed in a
discussion of the symmetry groups of the Gepner models, a classification of
subgroups consistent with supersymmetry, and the discussion of how orbifolding
by the maximal group can give rise to mirror models.

Secondly, we discussed the deep throat region of the non-compact Gepner models
and how to obtain the chiral primary states that are normalizable at weak
coupling from the conformal field theory. We then discussed Landau-Ginzburg
descriptions of both compact and non-compact Gepner models and extended the
existing techniques to analyze Landau-Ginzburg orbifolds such that they
applied to the non-compact models under discussion. This led to an intuitive
understanding of which modes become normalizable, and which modes are lifted
by a momentum or winding potential. The counting of deformations becomes very
tractable in the Landau-Ginzburg formalism. For completeness, we matched it
onto a more intricate conformal field theory counting.

We used these results to argue that mirror symmetry can be implemented in non-compact Gepner models. When taking into account all possible deformations of the linear dilaton theory, it becomes analogous to the compact case. However, one always needs to keep in mind that the choice of deformation needs to be mirrored when discussing the deformed theories.  As expected, we saw that in conformal field theory, mirror symmetry is implemented by a change in sign of the right-moving R-charge. The systematic treatment of symmetry groups allowed us to generate infinite classes of mirror pairs.

Indeed, in non-compact Gepner models, one can cancel off positive and negative contributions to the central charges of the individual factors, thus allowing for infinite classes of models that also have a small curvature limit. In such small curvature limits, we argued that one recuperates flat space with an
overall orbifold action. We identified such a limit, along with the orbifold, and showed that the conformal field theory matches with a flat space orbifold in the infinite level limit.  At finite level, we identified subtle differences in the spectrum of the conformal field theory and the toric abelian orbifold singularity. It would be interesting to find a general rule that tells us, a priori, which modes of the toric orbifold are retained in the conformal field theory. 

There are a large number of future directions that one can pursue.  For
instance, one can generalize the Landau-Ginzburg models to models with
fractional levels in the non-compact directions. One can also apply these
conformal field theory techniques to describe the spectrum of chiral primaries
and mirror theories for the heterotic string on non-compact Landau-Ginzburg
orbifolds.

Another direction that we had in mind while embarking upon this investigation
is the following. We have an orbifold approximation to particular non-compact
Gepner models.  Setting fractional or regular (i.e. physical) branes at such
toric orbifold singularities is one way to engineer interesting quiver gauge
theories. The toric data allows us to compute the superpotential on the brane
at the orbifold singularity using techniques that are very well developed.
Extending these results to determine the worldvolume superpotentials for
branes in non-compact Gepner models would be extremely interesting as such
results are not yet available using the exact boundary state description of
D-branes in these models.

Furthermore, Seiberg duality is well-understood in the context of the toric
quiver gauge theories \cite{Feng:2000mi, Beasley:2001zp, Cachazo:2001sg}. We
would like to study whether one can understand Seiberg duality for D-branes in
these almost toric spaces \cite{Eguchi:2003ik, Ashok:2005py,
  Fotopoulos:2005cn} microscopically, as in \cite{Murthy:2006xt,
  Ashok:2007sf}. The fact that we have a microscopic description of the
near-horizon doubly scaled limit of these backgrounds as well as a tunable
level, should give us further computational control.

\section*{Acknowledgments}
We are grateful to Agostino Butti, Eleonora Dell'Aquila, Bogdan Florea,
Davide Forcella, Amihay Hanany and Ruben Minasian for helpful conversations.
 Our work was supported in part
by the EU under the contract MRTN-CT-2004-005104. 

\appendix

\section{Non-compact Gepner model analysis}\label{CFTanalysis}

In this appendix we relate the counting of marginal deformations that we
performed
in a Landau-Ginzburg language to a more elaborate enumeration of states
in a standard conformal field theory formalism.
 See \cite{Eguchi:2004ik}
for a detailed discussion and further examples.

A state in the $(k_1,...,k_p;l_1,...,l_q)$ model (restricted to the internal
conformal field theory only) is associated to the left and right charges
$$r=(s_1,...,s_{p+q};n_1,...,n_p;2m_1,...,2m_q)$$
$$\tilde{r}=(\tilde{s}_1,...,\tilde{s}_{p+q};\tilde{n}_1,...,\tilde{n}_p;2\tilde{m}_1,...,2\tilde{m}_q)$$
as well as the compact spins $j_1,...,j_{p}$ and the non-compact spins
$j_{p+1},...,j_{p+q}$.  The spins are the same on the left and on the right,
since we consider diagonal partition functions. The $2 \beta_0$-orbifold
imposes integral R-charges, and allows the difference of left and right
charges $r-\tilde{r}$ to be an even multiple of the Gepner vector $\beta_0$.
The orbifolds $\beta_i$ that align the periodicities of the fermions allow
additional even differences between the left fermion numbers $s_i$ and the
right fermion numbers $\tilde{s}_i$.

Chiral primary operators are chiral primaries in each conformal field theory factor. From unitarity of the non-compact Gepner models (see appendix \ref{unitarity}) it  follows that in each factor separately we satisfy the equation $\pm Q_i = 2 h_i$ for chiral (respectively anti-chiral) primaries and similarly for the right-movers. Marginality of the deformations implies that we need to satisfy the equations $$ Q = 2 \beta_0\cdot r = \pm 1 \quad \tilde{Q} = 2 \beta_0\cdot \tilde{r} = \pm 1. $$

Solving this set of equations is straightforward but tedious because of the
equivalences that exist for the minimal model quantum numbers:
$$n_i \equiv n_i +2k_i,\quad s_i \equiv s_i+4,\quad (j,n,s) \equiv (\frac{k-2}{2}-j,n
-k,s+2)$$
The same kind of equivalences hold in the
non-compact factor:
$$2m_i \equiv 2m_i +2l_i,\quad s_i \equiv s_i+4,\quad (j,2m,s) \equiv (\frac{k+2}{2}-j,2m
-k,s+2)$$
 That is one technical reason why the Landau-Ginzurg
method is more efficient to count (anti)chiral operators.
However it is  possible to perform the counting in each individual example that we
treated in the bulk of the paper. We do this analysis example by example.

\subsection{The $(k;k)$ Model}
In the $(k;k)$ model, the Gepner vector is $\beta_0=(-1,-1;1;-1)$. Looking for
chiral primary operators, we find 
\begin{itemize}
\item $k-1$ $(c,c)$ states in the untwisted sector with charges:
$$ r=(0,0;n;k-n)=\tilde{r}, \quad j_1=n, \quad j_2=k-n, \quad 0 \le n \le k-2 $$
\item $k-1$ $(a,c)$ states in the first twisted sector:
$$ r=(0,0;-n;-k+n), \quad \tilde{r}=r-2\beta_0,\quad j_1=n, \quad j_2=k-n,  \quad 0 \le n \le k-2 $$
\item 1 $(c,c)$ state in each $\alpha$-twisted sector ($1 \le \alpha \le k-1$):
$$ r=(0,0;\alpha-1;k-\alpha+1), \quad \tilde{r}=r-2\alpha \beta_0,\quad
j_1=\alpha-1, \quad j_2=k-\alpha+1 $$
\item 1 $(a,c)$ state in each $(\alpha+1)$-twisted sector ($1 \le \alpha \le k-1$):
$$ r=(0,0;-\alpha+1;-k+\alpha-1), \quad \tilde{r}=r-2(\alpha+1) \beta_0, \quad
j_1=\alpha-1, \quad j_2=k-\alpha+1 $$
\end{itemize}
For each $(c,c)$ state we also find a $(a,a)$ state
and
each $(c,a)$ state is similarly paired with an $(a,c)$ state.
It is straightforward to match all these states with the ones we found more
fluently in the bulk of the paper with the Landau-Ginzburg methods. 

For each state, described by its quantum numbers $r$, $\tilde{r}$, $j_1$ and
$j_2$, we can write the corresponding closed string vertex
operator:
\be V_{r,\tilde{r}}^{j_1,j_2} = V^{j_1}_{n,s_1;\tilde{n},\tilde{s}_1}
V^{j_2}_{2m,s_2;2\tilde{m},\tilde{s}_2} \,. \ee
Here $V^{j_1}_{n,s_1;\tilde{n},\tilde{s}_1}$ is a vertex operator of
quantum numbers $(j_1,n,s_1)$ in the minimal model, 
while $V^{j_2}_{2m,s_2;2\tilde{m},\tilde{s}_2}$ is a vertex operator  
of quantum numbers $(j_2,n,s_2)$ in the noncompact factor. 
If we denote the asymptotic coordinates along the cylinder  as $\rho$ and
$\theta$
 and the bosonized complex fermion by $H$, we have the asympotic expression for the
 vertex
operators (see e.g. \cite{Hosomichi:2004ph}):
\be V^{j}_{2m,s;2\tilde{m},\tilde{s}} = \exp\left[ \sqrt{\frac{2}{k}}
  \left( (j-1)\rho +im \theta_L +i\tilde{m} \theta_R \right)
  +isH_L+i\tilde{s}H_R  \right] \,.\ee

\subsection{The $(2k,2k;k)$ Model}
\label{2k2kk}
Let us briefly mention the subtlety that the $\beta_0$ vector of Gepner is defined in
principle in all factors of the models. The reason we can consider
its action separately in the internal factors only without encountering
further difficulties lies in the fact that we first of all only consider even
multiples of $\beta_0$, and that moreover an even multiple of $\beta_0$ is
equivalent to the same vector with only non-zero internal entries, up to the
vectors $\beta_i$. That is true for all three-factor models we consider (and
it is therefore also true for the two-factor model when we incorporate two
further flat directions). 

Having dispensed of that subtlety in comparing the two methods, we
 can again find the marginal deformations directly in the non-compact Gepner model:
\begin{itemize}
\item The $(k-1)^2$ untwisted $(c,c)$ states have quantum numbers:
$$ r=(0,0,0;a,b;c)=\tilde{r}, \quad j_1=a, \quad j_2=b, \quad j_3=c$$
$$ 0 \le a,b \le 2k-2, \quad 2 \le c \le k, \quad a+b+2c=2k $$
\item The $k-1$ twisted $(a,c)$ states have quantum numbers:
$$r=(0,0,0;-2k+\alpha,-2k+\alpha;k-\alpha),\quad \tilde{r}=r-2\alpha \beta_0$$
$$j_1=2k-\alpha, \quad j_2=2k-\alpha, \quad j_3=\alpha-k, \quad k+2 \le \alpha \le 2k \,.$$
where $\alpha$ labels the twisted sector in which these states live.
\end{itemize}
\noindent
The corresponding closed string vertex operators are given by
\be V_{r,\tilde{r}}^{j_1,j_2,j_3} = V^{j_1}_{n_1,s_1;\tilde{n}_1,\tilde{s}_1}
V^{j_2}_{n_2,s_2;\tilde{n}_2,\tilde{s}_2}
V^{j_3}_{2m,s_3;2\tilde{m},\tilde{s}_3}\,.\ee

\subsection{The $(k; 2k, 2k)$ Model}\label{k2k2k}

The marginal deformations are written as follows:
\begin{itemize}
\item The $(k-1)^2$ untwisted $(c,c)$ states have quantum numbers:
$$ r=(0,0,0;a;b,c)=\tilde{r}, \quad j_1=a, \quad j_2=b, \quad j_3=c $$
$$ 0 \le a \le k-2, \quad 2 \le b,c \le 2k, \quad 2a+b+c=2k $$
\item The $k-1$ twisted $(a,c)$ states have quantum numbers:
$$ r=(0,0,0;-k+\alpha;-\alpha,-\alpha), \quad \tilde{r}=r-2\alpha \beta_0$$
$$j_1=k-\alpha, \quad j_2=\alpha, \quad j_3=\alpha, \quad 2 \le \alpha \le k $$
where $\alpha$ labels the twisted sectors as before.
\end{itemize}
\noindent
The corresponding closed string vertex operators are given by:
\be V_{r,\tilde{r}}^{j_1,j_2,j_3} = V^{j_1}_{n,s_1;\tilde{n},\tilde{s}_1}
V^{j_2}_{2m_1,s_2;2\tilde{m}_1,\tilde{s}_2}
V^{j_3}_{2m_2,s_3;2\tilde{m}_2,\tilde{s}_3}\,.\ee

For the orbifold models as well, one can perform the tedious exercise, thus
affirming that the Landau-Ginzburg formalism is indeed more efficient.

\subsection{The $(3,3,3;2)$ Model}\label{3332}

There is an extra subtlety that we need to address for this model. From the
conformal field theory point of view, it is easiest to 
ignore the quadratic Landau-Ginzburg model with a trivial infra-red fixed
point,
and to work with an even number of internal conformal field theory factors. It then follows that the vector 
\begin{equation}
2\beta_0=(-2,-2,-2,-2,-2;2,2,2;-2)
\end{equation}
in the full light-cone conformal field theory is not equivalent modulo the
vectors $\beta_i$ to the vector $2 \gamma_0 =(0,0,0,0,0;2,2,2;-2)$ which
represents the $g_0$ action on the Landau-Ginzburg internal conformal field
theory. Yet, it can be show that the Landau-Ginzburg model does correctly
count the conformal field theory chiral-chiral and chiral-anti-chiral states.
Since it is only the $(c,a)$ state in the $3$-twisted sector that is crucial
to us in this example, let us show how to identify that state in the conformal
field theory. It corresponds to the state with charges $r=(0,0,0,0,0;0,0;2)$
on the left, as indicated by the Landau-Ginzburg model. The right-moving
charges are computed by observing that we are in the $3$-twisted sector. We
obtain the charge (up to equivalences in the charge lattice)
$\tilde{r}=(0,0,0,0,-2;0,0;0)$ (and we remain diagonal in the non-compact
quantum number $j=1$). Indeed, the state with these charges is in the spectrum
of the theory.

Note that we had to adjust the precise identification of the state in the conformal field theory. This is typical of the model with an even number of factors. The final vertex operator is chiral and bosonic on the left, and anti-chiral and purely made of fermions on the right.

\section{T-duality to NS5-branes revisited}\label{NS5}
In this appendix, we recall the relation of the $(SU(2)/U(1) \times
SL(2,R)/U(1))/\IZ_k$ coset model to the near-horizon geometry of a particular
constellation of NS5-branes \cite{Ooguri:1995wj, Sfetsos:1998xd, Giveon:1999px}. It is useful to revisit this exercise because we will be able to explicitly identify the region of
space-time in which the NS5-branes reside with the presence of a
patch isomorphic to $\,\IC^2/\IZ_k$ in the T-dual. This fact is used as an
argument in 
section \ref{geometrymirror}.

\begin{figure}
\centering
\includegraphics[width=0.8\textwidth]{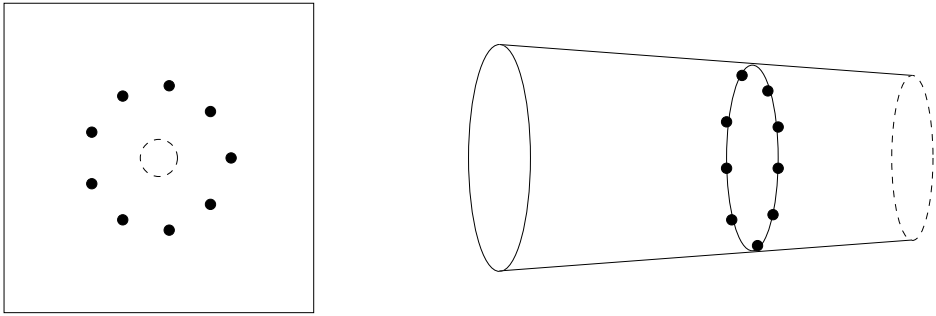}
\caption{NS5-branes spread on a topologically trivial circle (drawn on the left). 
When we cut out a little disc at the center, the configuration becomes topologically equivalent to NS5-branes spread on a topologically non-trivial circle (drawn on the right).}
\label{planeALF}
\end{figure}

We recall that the supergravity background generated by
parallel NS5-branes stretching in
 the $x^{\mu=0,1,2,3,4,5}$-directions in the string frame is:
\begin{eqnarray}
ds^2 &=& \eta_{\mu\nu} dx^{\mu} dx^{\nu} + H(x^i) dx^i dx_i \nonumber \\
e^{2 \Phi}  &=& g_s^2 \, H(x^i) \nonumber \\
H_{ijk} &=& - {\epsilon^l}_{ijk} \partial_l H(x^i)
\end{eqnarray}
where the harmonic function $H$ is determined by the positions of
the NS5-branes $x^{i=6,7,8,9}_a$:
\begin{eqnarray}
H(x^i) &=& 1 + \sum_{a=1}^k \frac{\alpha'}{|x^i-x^i_a|^2}
\end{eqnarray}
We concentrate on $k$ NS5-branes spread evenly
on a topologically trivial circle of coordinate 
radius $\rho_0$ in the $(x^6,x^7)$
plane (see figure \ref{planeALF}). We recall from 
\cite{Sfetsos:1998xd, Israel:2005fn} that after the coordinate
change $r \ge 0$ and $\theta \in {[} 0 , \pi/2 {]} $
\begin{eqnarray}
(x^6,x^7) &=& \rho_0 \cosh r \sin \theta \, (\cos \psi,\sin \psi)  \nonumber \\
(x^8,x^9) &=& \rho_0 \sinh r \cos \theta \, (\cos \phi,\sin \phi)\,
\label{cartcoords}
\end{eqnarray}
and taking a near-horizon doubly scaled limit in which $\rho_0 \rightarrow 0$
and $g_s \rightarrow 0$ and in which
 $\rho_0 g_s / \sqrt{\alpha'}$ (and $\alpha'$) is kept fixed \cite{Giveon:1999px},
and after neglecting the localization of the NS5-branes on the circle,
we obtain the NS5-brane background:
\begin{eqnarray}
ds^2 &=& dx^{\mu} dx_{\mu} + \alpha' k \,
\left[ dr^2+d \theta^2 + \frac{\tanh^2 r \ d \phi^2 + \tan^2 \theta \ d \psi^2
}{1+\tan^2 \theta \tanh^2 r} \right] \, ,
\nonumber \\
e^{ 2 \Phi} &=& \frac{g_\textsc{eff}^2 }{\cosh^2 r - \sin^2 \theta} \, ,
 \nonumber \\
B  &=& \frac{\alpha' k }{1+\tan^2 \theta \tanh^2 r} \ d \phi \wedge d \psi \,
\label{tdualNS5geom}
\end{eqnarray}
where the effective string coupling constant is 
\begin{equation} 
g_\textsc{eff} = \frac{\sqrt{k \alpha'}g_s}{\rho_0}. 
\end{equation}
We refer to \cite{Israel:2005fn} for the detailed calculation.

A first observation to make is that the NS5-branes are located
at $r=0$ and $\theta=\pi/2$. Moreover, the fact that the coordinate
radius of the NS5-brane ring has gone to zero, has been compensated by the
fact that the harmonic function and radial metric
component has blown up 
and in such a way that the interior of the NS5-brane ring still has a
proper size of order $\sqrt{k \alpha'}$. It is difficult to press together
NS5-branes. 
That's an important aspect of the solution, since in the interior, near $r=0$
and $\theta=0$, we simply find a portion of flat space.

After T-duality in the angular direction $\psi$ around which
the NS5-branes have been sprinkled, we found the T-dual geometry:
\begin{eqnarray}
ds^2 &=& \alpha' k \left[ d r^2 + \tanh^2 r \, \left( \frac{d\chi}{k}\right)^2
+ d\theta^2 + \mathrm{cotan}^2 \theta \, \left( \frac{d\chi}{k}-d\phi \right)^2 \right], \nonumber \\
e^{2 \Psi}&=& \frac{g_\textsc{eff}}{k} \ \frac{1}{\cosh^2 r \ \sin^2 \theta}
\end{eqnarray}
where $\chi$ is an angle parameterizing the T-dual circle.
The background is recognized as a vector $\IZ_k$ orbifold of the 
product of coset conformal field theory geometries
$SU(2)_k /U(1) \, \times SL(2,R)_k /U(1)$.

We want to add a couple of remarks to the analysis in \cite{Israel:2005fn}
 to which
we refer the reader for further comments. In
particular, we first note the  singularity in both the metric
and the dilaton at $\theta=0$. The origin of this singularity is the fact that
we have performed an angular T-duality with a fixed point. Around the fixed
point, as we have pointed out, the original background behaves like flat
space. Thus, the T-dual behavior is recognized as the same type of
singularity that one obtains in T-dualizing flat space in cylindrical
coordinates. Since the original theory was regular near the origin, we expect the T-dual
theory to behave well at this point as well. 

On the other hand, we observe that the region at $\theta= \pi/2$ and $r=0$
where the NS5-branes resides has locally become identical to a $\,\IC^2/\IZ_k$
orbifold. Thus, it is the orbifold singularity of order $k$ that codes the
presence of the NS5-branes in the T-dual. 

We can make the above analysis  more precise by performing the
following mental exercise. Cut out from the $6-7$ plane a little disc at the
origin. Topologically, the space transverse to the NS5-branes has become
$\IR^3 \times S^1$, with NS5-branes spread on the $S^1$ (see figure \ref{planeALF}).
When we neglect the localization of the NS5-branes on the circle, the configuration is T-dual to
an ALF space \cite{Ooguri:1995wj}\cite{Kutasov:1995te}, which in the particular case where the
NS5-branes coincide in $\IR^3$ develops a $\,\IC^2/\IZ_k$ orbifold singularity. 
That reasoning is another version of the one above, which uses local
 fiberwise T-duality. Localizing the NS5-branes in
 the original geometry on the circle is known to be equivalent to taking into
 account worldsheet instanton corrections for the ALF space  \cite{Gregory:1997te}\cite{Tong:2002rq}\cite{Okuyama:2005gx}, and this
 equivalence is conjectured to be valid for the NS5-branes on a topologically
 trivial circle as well \cite{Israel:2005fn}.

We take away several facts from this analysis. Firstly, the $\IZ_k$
orbifold projection acting on the  $SU(2)/U(1) \times
SL(2,R)/U(1)$ coset has fixed points and gives rise locally
to a physical $\IC^2/\IZ_k$ singularity that is T-dual to the presence of 
$k$ NS5-branes. 
And secondly, 
in the geometry of the coset conformal field theory, the $\IZ_k$ GSO
projection acts geometrically on the cigar coordinates,
and this in contrast to the fact that the GSO projection in Gepner
models does not orbifold the coordinates
 of weighted projective space. We recalled
the above detailed example because we will use 
these facts as arguments in section \ref{geometrymirror} and they are particularly
manifest in the above example.

\section{The consequences of unitarity}\label{unitarity}

We list in this appendix a number of properties that hold in unitary modules
of the $N=2$ superconformal algebra, irrespective of the value of the central
charge (and in particular also when the central charge is equal or greater
than three). Our conventions for the $N=2$ superconformal algebra coincide
with those of \cite{Polchinski:1998rr, Lerche:1989uy}.
\subsection*{Properties}
The following properties hold -- the proofs in the references that
we give depends only on the unitarity of the 
representation spaces as we checked on a case-by-case basis:
\begin{enumerate}
\item All states in the NS-sector with conformal dimension $h$ and R-charge
$Q$ satisfy the inequality $h \ge |Q|/2$ \cite{Lerche:1989uy}.
\item
An NS-sector field of conformal dimension $h$ and R-charge $Q$ is a
chiral primary 
if and only if $h=Q/2$ and anti-chiral if and only if $h=-Q/2$
\cite{Lerche:1989uy}. 
\item Chiral primary fields satisfy the inequality $h \le c/6$ \cite{Lerche:1989uy}.
\item Ramond sector ground states have R-charges $Q$ in the range
$ -c/6 \le Q \le +c/6$ \cite{Lerche:1989uy}.
\item An NS-sector state $|\phi \rangle$ 
satisfies the equations 
$G^+_{-l-1/2} | \phi \rangle = 0 = G^-_{l+1/2} |\phi \rangle$
if and only if its conformal dimension
 $h$ and R-charge $Q$ satisfy the relation $h=(l+1/2) Q-c(l^2+l)/6$ \cite{Vafa:1989xc}.
\item A chiral ring exists \cite{Lerche:1989uy}.
\end{enumerate}
\subsection*{Remarks}
Note that we performed a non-trivial exercise. For instance, for conformal
field theories with central charge $c<3$, it is argued in \cite{Lerche:1989uy}
 that there
always exist a chiral primary field in the conformal field theory
with conformal dimension $h=c/6$. That is a consequence of the existence of 
the unit operator in the theory (combined with spectral flow).
 In other words, there is a normalizable
$SL(2,R)$ invariant ground state in these conformal field theories. That is
not  a consequence of unitarity only, and it does {\em not}
 hold generically for unitary $N=2$ superconformal field theories. In fact, the situation is
subtle. For instance there are examples of bulk $N=2$ 
superconformal field theories with central charge $c>3$ for which the unit
operator does not exist when the conformal field theory is
defined on a Riemann surface without boundary,
while it does exist for instance on the boundary of a
disc with particular boundary
conditions (see e.g. \cite{Eguchi:2003ik}\cite{Israel:2004jt}). 
It is due to these kind of subtleties that it is useful to have a
(partial) list of properties that we can indiscriminantly use for unitary 
$N=2$ theories with any central charge.

\end{document}